\documentclass{WileyMSP-template}

\usepackage[english]{babel}
\usepackage{lipsum}  
\usepackage{amsmath}
\usepackage{graphicx}
\usepackage{svg}
\usepackage{tabularx,multirow}
\usepackage{longtable,threeparttablex}
\graphicspath{{figs/}}
\usepackage{hyperref}
\hypersetup{
 unicode=false,			
 pdftoolbar=true,		
 pdffitwindow=false,		
 pdfstartview={FitH},		
 pdfcreator={pdflatex},		
 pdfnewwindow=true,		
 colorlinks=true,		
 linktoc=page,			
 linkcolor=blue,		
 citecolor=blue,		
 filecolor=blue,		
 urlcolor=blue			
 }
\usepackage{soul}
\usepackage[normalem]{ulem}

\DeclareUnicodeCharacter{2212}{-}

\usepackage[backend=biber, style=numeric, sorting=none]{biblatex}
\usepackage{csquotes}

\begin{document}

\pagestyle{fancy}
\rhead{\includegraphics[width=2.5cm]{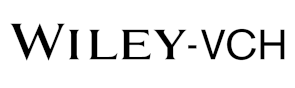}}

\title{Coherent control of nitrogen nuclear spins via the V$_B^-$-center in hexagonal boron nitride}

\maketitle


\author{Adalbert Tibiássy, Charlie J. Patrickson, Thomas Poirier, James H. Edgar, Bruno Lopez-Rodriguez, Viktor Ivády, Isaac J. Luxmoore}

\vspace{0.4 cm}

\begin{affiliations}
Adalbert Tibiássy,  Viktor Ivády \\
\emph{Department of Physics of Complex Systems, E\"otv\"os Lor\'{a}nd University,\\
Egyetem t\'{e}r 1-3, H-1053 Budapest, Hungary, and } \\
\emph{MTA–ELTE Lend\"{u}let "Momentum" NewQubit Research Group,\\
P\'{a}zm\'{a}ny P\'{e}ter, S\'{e}t\'{a}ny 1/A, 1117 Budapest, Hungary}\\

\vspace{0.4 cm}

Charlie J. Patrickson and Isaac J. Luxmoore \\
\emph{Department of Engineering, University of Exeter, EX4 4QF, United Kingdom}\\

\vspace{0.4 cm}

Thomas Poirier and James H. Edgar \\
\emph{Tim Taylor Department of Chemical Engineering, Kansas State University, Manhattan, Kansas 66506, USA}\\

\vspace{0.4 cm}

Bruno Lopez-Rodriguez\\
\emph{Department of Imaging Physics (ImPhys), Faculty of Applied Sciences, Delft University of Technology, 2628 CJ Delft, The Netherlands}\\

\vspace{0.4 cm}

Email Address: ivady.viktor@ttk.elte.hu and i.j.luxmoore@exeter.ac.uk

\end{affiliations}



\justifying






\vspace{1cm}

\newpage

\begin{abstract}
\noindent
\large{Charged boron vacancies (V$_\text{B}^-$) in hexagonal boron nitride (hBN) have emerged as a promising platform for quantum nanoscale sensing and imaging. While these primarily involve electron spins, nuclear spins provide an additional resource for quantum operations. This work presents a comprehensive experimental and theoretical study of the properties and coherent control of the nearest-neighbor $^{15}$N nuclear spins of V$_\text{B}^-$-ensembles in isotope-enriched h$^{10}$B$^{15}$N. Multi-nuclear spin states are selectively addressed, enabled by state-specific nuclear spin transitions arising from spin-state mixing. We perform Rabi driving between selected state pairs, define elementary quantum gates, and measure longer than 10~$\mu$s nuclear Rabi coherence times. We observe a two orders of magnitude nuclear g-factor enhancement that underpins fast nuclear spin gates. Accompanying numerical simulations provide a deep insight into the underlying mechanisms. These results establish the foundations for leveraging nuclear spins in V$_\text{B}^-$ center-based quantum applications, particularly for extending coherence times and enhancing the sensitivity of 2D quantum sensing foils.
}
\end{abstract}

\newpage

\section*{Introduction}

Spin defects in wide band-gap semiconductors have emerged as a promising platform for nanoscale quantum sensing, dominated by the nitrogen vacancy (NV) center in diamond, thanks to its long coherence times and efficient optical spin polarization at room temperature. NV centers can be applied in dense ensembles to reach $\mathrm{pT/\sqrt(Hz)}$ level sensitivities \cite{Fescenko2020,Wang_SciAdv_2022}, or mounted individually in diamond tips to probe magnetism with nanoscale spatial resolution \cite{Rondin_2014,Xu2023}. There is growing interest in directly integrating quantum sensors with the probed system to achieve higher spatial resolution. Examples include introducing spin defects in nanoscale particles to biological samples \cite{schirhagl_nitrogen-vacancy_2014,Wu2022,Wang2023}, and direct integration of the system of interest with a quantum sensing chip \cite{Briegel2025,Cambria2025,Cheng2025}. This is challenging with NV centers, as the diamond host is difficult to process, and the NV coherence time is substantially degraded when located close to the sample surface due to charge trapping effects. In contrast, with hexagonal boron nitride (hBN), the defects remain effective when they are close to the material surface. Consequently, they can be in proximate contact or be embedded within the target system, and are inherently well-suited to heterogeneous integration thanks to well-developed pick and place fabrication methods. This approach has been applied to study magnetism in 2D \cite{Huang2022,Kumar2022,healey_quantum_2021} and bulk \cite{Zhou2024,Zang2024} magnetic materials, with ensembles of boron vacancy defects $V_B^-$ in hBN flakes directly integrated with the target platforms.  

The $V_B^-$ has been thoroughly studied by experiments \cite{gottscholl_initialization_2020} and theory \cite{ivady_ab_2020,sajid_edge_2020}, testifying the keen interest and potential of spin defects in 2D materials. Like the negatively charged nitrogen vacancy (NV) center in diamond, $V_B^-$ is a radiative spin triplet system (total spin quantum number, S = 1). It has ground- and excited-state zero-field splitting of $\sim$3.5 GHz\cite{gottscholl_initialization_2020} and $\sim$2.1 GHz\cite{baber_excited_2022,mathur_excited-state_2022}, respectively. Spin initialization can be achieved through optical pumping. Room-temperature electron spin relaxation, $T_1$ and spin coherence, $T_2$, times are$\sim20~\mathrm{\mu s}$ \cite{Gottscholl_SciAdv_2021} and $100~\mathrm{ns}$ \cite{haykal_decoherence_2022,Ramsay2023,Rizzato2023, tarkanyi_understanding_2025, lee_magnetic-field_2025}, respectively. Various dynamical decoupling schemes can effectively extend $T_2$ \cite{Ramsay2023,Rizzato2023,Gong2023}, and enhance sensing performance \cite{Rizzato2023,Patrickson2024,Patrickson2025}. Consequently, the $V_B^-$ center is an alternative to the NV center due to its potentially superior performance in near-surface sensing and imaging applications.\cite{daly_prospects_2025}

Strong hyperfine coupling provides a mechanism to polarize and control the three nearest neighbor nitrogen nuclei via the $V_B^-$ electron spin \cite{gao_nuclear_2022,liu_coherent_2022,gong_isotope_2024,CluaProvost2023,Ru2024,mamin_probing_2025}.

Optical pumping via spin mixing near the avoided crossings of the excited and ground states (ESLAC and GSLAC) has been shown to partially polarize the nitrogen nuclei \cite{gao_nuclear_2022,gong_isotope_2024,CluaProvost2023,Ru2024}, while coherent control of the nuclear spins can be achieved using electron-nuclear double resonance (ENDOR) techniques \cite{gao_nuclear_2022,gong_isotope_2024,mamin_probing_2025}.

In this article, we explore and demonstrate the coherent control of $^{15}$N nitrogen nuclear spins adjacent to the V$_\text{B}^-$ center in an isotopically enriched hBN crystal~\cite{janzen_boron_2024}. We perform an electron-nuclear double resonance study and selectively address the four nuclear spin states of the three identical spin-1/2 nitrogen nuclei. Theory predicts that this selective addressability arises from electron-nuclear hyperfine coupling, which induces state-dependent mixing of the two spin species. This mixing results in distinct shifts lifting the degeneracy of the transitions. Hyperfine mixing significantly enhances the nuclear g-factor, resulting in an increase of the Rabi frequency by approximately a factor of 116 compared to that of pure nuclear spin states. In addition, Rabi oscillation of the nuclear spin has a coherence time exceeding 10~$\mu s$ and there is an unexpected loss of nuclear polarization when pumping the defect optically.  The narrow magnetic resonance line width, long coherence time, and fast nuclear spin control make nuclear spins adjacent to V$_\text{B}^-$ centers a desirable platform for quantum applications, especially in the field of quantum sensing.

\begin{figure*}
\includegraphics[width=1.0\textwidth]{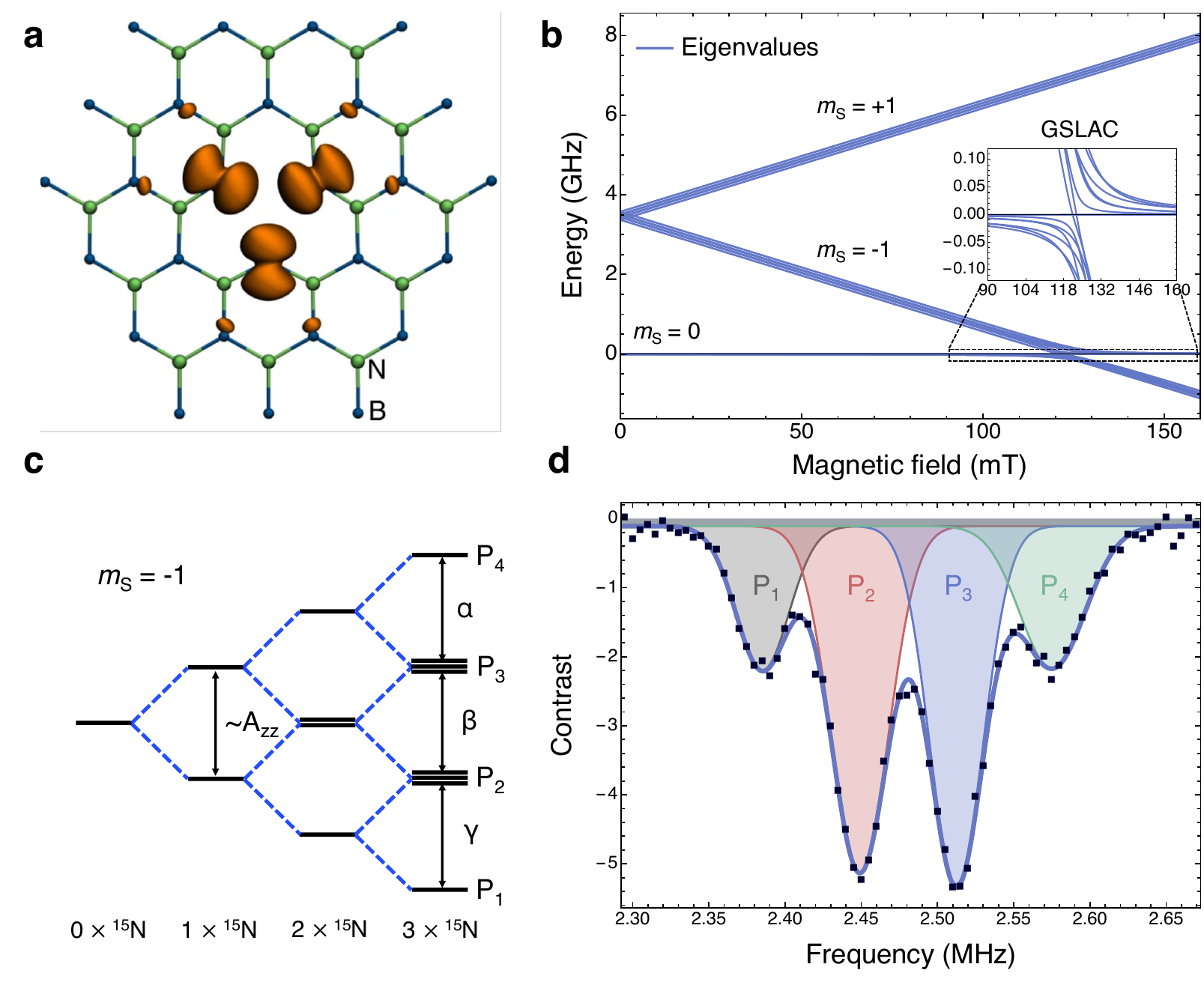}
	\caption{ \textbf{Configuration and fine electronic structure of the V$_\text{B}^-$ center in h$^{10}$B$^{15}$N.} \textbf{a} Structure and electron spin density (dark orange lobes) of the V$_\text{B}^-$-center in hBN. Green and deep blue spheres depict nitrogen and boron atomic positions. \textbf{b} Magnetic field dependence of the spin energy levels. The inset depicts the energy levels at the ground state level avoided crossing  (GSLAC) region of strong electron-nuclear spin mixing. \textbf{c} Emergence of the hyperfine structure in the $m_{\text{S}} = -1$ branch in hB$^{15}$N away from the GSLAC. \textbf{d} ODMR spectra of the V$_\text{B}^-$ center in h$^{10}$B$^{15}$N at 212~mT fitted with a sum of four Gaussian curves. }
	\label{fig:fig_struc}  
\end{figure*}

Before presenting our results, we briefly review the most relevant properties of the V$_\text{B}^-$ center that provide the basis for our discussions. The V$_\text{B}^-$ center is a planar defect in hBN with D$_{3\text{h}}$ symmetry. The defect electronic states in the bandgap arise from in-plane nitrogen dangling bonds, which define the spin density (see Fig.~\ref{fig:fig_struc}a), and from the out-of-plane $p_z$ orbitals of the neighboring nitrogen atoms. The latter orbitals play an important role in forming the optically excited triplet state \cite{ivady_ab_2020,reimers_photoluminescence_2020}. Each lattice site in hBN  contains a nuclear magnetic moment that interacts with the electron spin through dipole-dipole and Fermi contact hyperfine interaction terms obtained from the spin density distribution of the defect. In our experiments and simulations, we study boron-10 and nitrogen-15 isotope enriched $\mathrm{h^{10}B^{15}N}$ crystals. Due to the localized nature of the defect orbitals and thus the spin density, see Fig.~\ref{fig:fig_struc}, the V$_\text{B}^-$ center couples most strongly to the three nearest-neighbor $^{15}$N nuclear spins ($I=1/2$). Indeed, the dynamics of the V$_\text{B}^-$ electron spin cannot be considered separately from its first neighbor nuclear spins, and these four spins form an entangled core.\cite{liu_coherent_2022,cholsuk_spin_2025}  The second and third-strongest hyperfine interactions are approximately one and two orders of magnitude smaller\cite{ivady_ab_2020,cholsuk_spin_2025}, respectively, allowing us to investigate the properties of the electron spin and its three nearest-neighbor nuclei independently of the surrounding nuclear spin bath in first order. 

Hyperfine interaction with the nearest neighbor $^{15}$N spin is crucially important in our study, thus, we examine different terms of the coupling and their effect in the presence of an external magnetic field separately.  In this study, we employ a constant $B$ magnetic field  parallel to the $c$-axis, which is also aligned with the preferential quantization axis of the defects set by the zero-field splitting interaction of the ground-state electron spin. The static Hamiltonian for the electron and the three $^{15}$N nuclear spin can be written as,
\begin{equation} 
        \hat{H} = D S_z^2 + \gamma_e S_z B  +\sum_i^3 \left( \gamma_{^{15}N} I_{z,i} B +  \mathbf{S} A_i \mathbf{I}_{i} \right), 
\end{equation}
where $\gamma_e$ and $\gamma_{^{15}N}$ are the gyromagnetic ratios of the electron and the nuclear spins, $\mathbf{S}$ and $\mathbf{I}_{i}$ are the electron and nuclear spin operator vectors, respectively, and $D$ is the parallel zero-field splitting parameter. The transverse zero field splitting parameter $E$ is neglected in our study, since we focus predominantly on the 150-250~mT magnetic field range, where $E$ has a negligible effect. Finally, $A_i$ are the hyperfine coupling tensors of the first neighbor nuclear spins, for which we initially use published theoretical tensors\cite{ivady_ab_2020} and later scale to perfectly match our experimental nuclear magnetic resonance spectra.

Magnetic field dependence of the spin eigenstates of the four-spin system is depicted in Fig.~\ref{fig:fig_struc}b. The eigenstates are separated into three branches according to the electron spin $m_{\text{S}}$ quantum numbers, except at the ground state level avoided crossing (GSLAC) where the $m_{\text{S}} = 0$ and the  $m_{\text{S}} = -1$ states are heavily mixed by the hyperfine coupling of the first neighbor nuclei $^{15}$N, see Fig.~\ref{fig:fig_struc}(b). Away from the GSLAC, the secular hyperfine coupling $ A_\parallel = A_{zz} $ of the three first neighbor spin-1/2 $^{15}$N nuclear spins gives rise to a characteristic four-peak hyperfine structure\cite{CluaProvost2023} of the $m_{\text{S}} = \pm1$ branches of the V$_\text{B}^-$ center with 1-3-3-1 degeneracies in h$^{10}$B$^{15}$N, see Fig.~\ref{fig:fig_struc}c. Fig.~\ref{fig:fig_struc}c details how the degenerate states are formed, assuming each added nuclear spin splits the energy levels equally (see the dashed blue lines), corresponding to the secular $A_\text{zz}$  for each nucleus. Since the system exhibits a 120$^\circ$ rotation symmetry, all first neighbor $^{15}$N $A_\text{zz}$ components are the same.

\section*{Results}

First, we carry out optically detected magnetic resonance (ODMR) on the ensemble of V$_\text{B}^-$ in h$^{10}$B$^{15}$N by driving the $m_S = 0 \leftrightarrow m_S = -1$ transition at 212~mT. A characteristic four-peak hyperfine structure is observed, see Fig.~\ref{fig:fig_struc}d, where the peaks are labeled as P$_1$-P$_4$ in order of increasing energy for the $m_{\text{S}} = -1$ state, and correspond to $m_{Jz} = \left \lbrace +3/2, +1/2, -1/2, -3/2 \right\rbrace $ quantum numbers of $J_z = \sum_i I_{z,i}$ in the secular approximation, i.e., when only the diagonal elements of the spin Hamiltonian are considered. 
When the nuclear spins are not polarized, i.e.\ all states are occupied with equal probability, the areas under the ODMR peaks P$_1$-P$_4$ are proportional to the  1-3-3-1 degeneracy of the hyperfine levels.

\begin{figure*}
\centering
\includegraphics[width=1.0\textwidth]{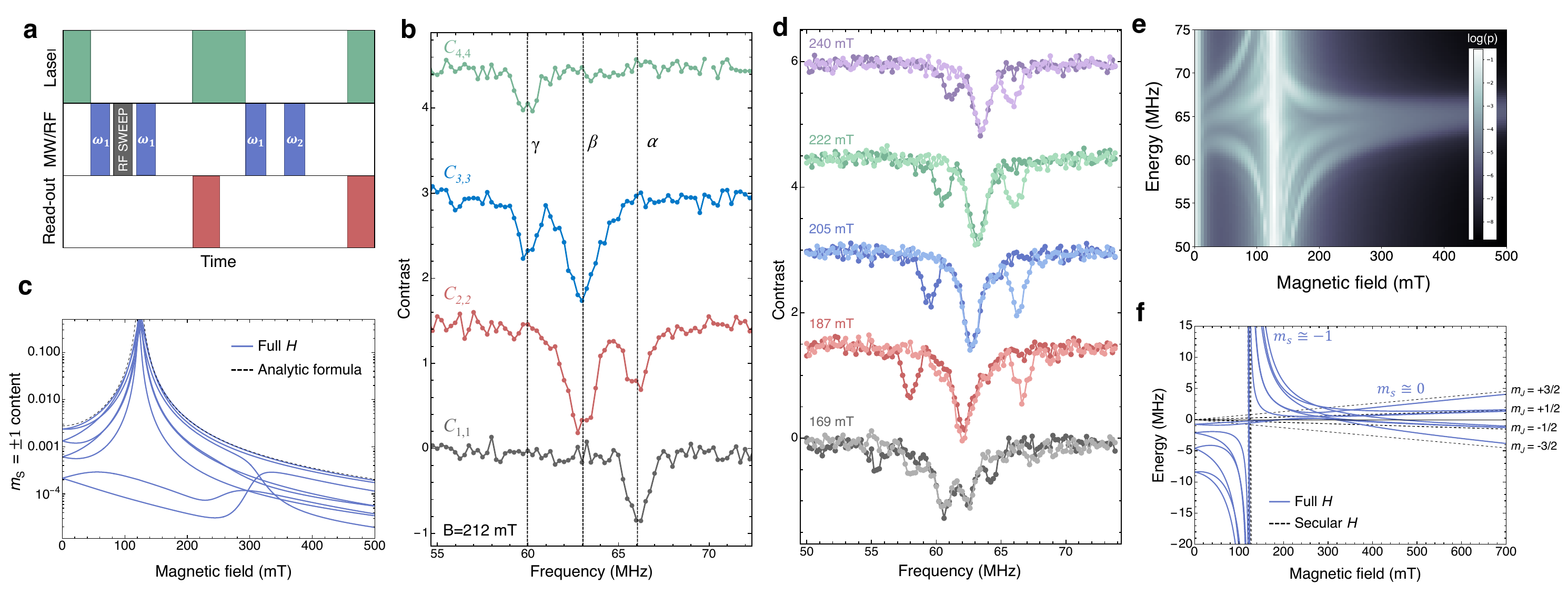}
	\caption{ \textbf{Electron nuclear double resonance of first neighbor nitrogen nuclear spins.} \textbf{a}  Pulse sequence of nuclear spin state control. Sage green, slate blue, and gray rectangles represent laser, microwave (MW), and radio frequency (RF) pulses, while the  terracotta pink rectangles represent the period of photon collection. The pulse durations are not to scale.  \textbf{b}  Electron nuclear double resonance spectra (ENDOR) the V$_\text{B}^-$ center in  h$^{10}$B$^{15}$N. Green, blue, terracotta, and gray curves show the ENDOR contrast $C_{i,i}$ obtained by initializing and measuring the same $P_i$ hyperfine sublevel, see Fig.~\ref{fig:fig_struc}d. Here, an external magnetic field of 212~mT is used in the measurements. \textbf{c} The amount of mixing of the electron spin $m_{\text{S}} = \pm1$ states with the dominantly $m_{\text{S}} = 0$ state. The value 0.01 means 1\% mixing. Blue lines depict the mixing of the hyperfine resolved states obtained by diagonalizing the full Hamiltonian, while the dashed gray line depicts the result of a simple analytical formula assuming only electron spin state splitting and hyperfine mixing terms of strength $A_\perp$, see text.
    \textbf{d} Magnetic field dependence of the ENDOR resonance peaks. \textbf{e} simulated wide field magnetic dependence of the ENDOR resonance peaks. The density map $p$ is obtained by summing up all $C_{i,i}$ curves at all magnetic field values. \textbf{f} Close-up view of the magnetic field dependence of the energy levels of the $m_{\text{S}} = 0$ branch. }
	\label{fig:fig_shift}  
\end{figure*}

To selectively address the nuclear spin states of the  V$_\text{B}^-$ center, we carry out pulsed electron-nuclear double resonance studies on our isotope-enriched sample, see Methods for further details. We first initialize the electron spin predominantly in the $m_{\text{S}} = 0$ subspace by a blue (488~nm) laser pulse and apply a microwave (MW) $\pi$-pulse with frequency $\omega_i$ (CNOT  gate) to selectively populate one of the peaks P$_1$-P$_4$. The population transferred by the MW pulse is proportional to the lowest degeneracy of the hyperfine levels involved in the transitions. After some population is transferred to the $m_{\text{S}} = -1$ branch, a radiofrequency (RF) pulse with a varying frequency is applied to drive resonant transitions between the populated and unpopulated hyperfine states. The duration of the pulse is set to approximately perform a $\pi$-pulse between the hyperfine states when the RF drive is resonant with the transition. Finally, another MW  $\pi$-pulse with frequency $\omega_j$ is applied to selectively test the population of peak P$_j$. After performing this pulse sequence, the electron spin population carries information on the success of the nuclear spin transition, which is then read out via the photoluminescence intensity at the beginning of the next laser pulse. The sequence is repeated without applying the RF pulse for better contrast. Finally, the pulsed ENDOR signal is obtained as 
\begin{equation}
    C_{i,j} = \frac{ I_{\text{on}}^{i,j} - I_{\text{off}}^{i,j} }{I_{\text{off}}^{i,j}},
\end{equation}
where $I_{\text{on (off)}}^{i,j}$ is the PL intensity detected while addressing peaks P$_i$ and  P$_j$ with MW transition frequencies $\omega_i$ and $\omega_j$ for RF drive on (off).

A sample ENDOR spectrum is depicted in Fig.~\ref{fig:fig_shift}b. In each measurement, we populate peaks P$_1$, P$_2$, P$_3$, or P$_4$ with the first MW pulse and examine the population of the same peaks after the RF pulse ($C_{i,i}$ signals). Dips in the contrast indicate successful nuclear spin population transfer from the initial hyperfine state $i$ to other hyperfine states with $\Delta m_{J,z} \simeq \pm1$. As can be seen in Fig.~\ref{fig:fig_shift}b, starting from the lowest energy P$_1$ state, there is only one nuclear resonance transition. For the case of middle P$_2$ and P$_3$ states, there are two possible transition frequencies, while for the P$_4$ state, again, a single transition energy is revealed by sweeping the frequency of the RF pulse. Altogether, there are three distinct transition energies, $\alpha$, $\beta$, and $\gamma$ are observed, see Fig.~\ref{fig:fig_struc}c and Fig.~\ref{fig:fig_shift}b. The dip amplitudes are proportional to the lowest degeneracies of the states involved in the transition. Our protocol thus enables selective initialization of the 4-spin system in all the hyperfine states in the $m_{\text{S}} = -1$ branch.  Finally, we draw attention to the line width of the ENDOR curves. The nuclear resonance dips exhibit linewidths as narrow as 1.2~MHz (FWHM), which may make these features useful for sensing.

Strikingly, the $\alpha$, $\beta$, and $\gamma$ transitions exhibit notable inequivalence, even though the secular $A_\parallel$ components of the nitrogen hyperfine tensors, which determine the transition energies in first order, are equivalent under threefold rotational symmetry. However, the significant perpendicular components, $\left| A_{\perp} \right| \sim 100$~MHz, give rise to mixing between the different $m_S$ electron spin states. This mixing induces nuclear-spin-state-dependent shifts that, in turn, lead to measurable differences in the $\alpha$, $\beta$, and $\gamma$ transition energies. Moreover, since the degree of mixing varies across the investigated range of external magnetic fields (Fig.~\ref{fig:fig_shift}d), the $\alpha$-$\gamma$ transition energies consequently exhibit magnetic-field dependence. A comprehensive discussion and a perturbation-theory analysis of the mixing and the resulting energy shifts are provided in Supplementary Note~2.

To test our theory, we perform pulsed ENDOR measurements at various magnetic field values as depicted in Fig.~\ref{fig:fig_shift}d. The position of the nuclear resonance dips shows considerable magnetic field dependence in the examined 169-240~mT magnetic field interval, in agreement with the theoretical expectations. Closer to the GSLAC, we observe not only the shift of the $\alpha$-$\gamma$ transition energies but also the splitting of the $\beta$ transition. 

The measurement in Fig.~\ref{fig:fig_shift}d enables us to determine the hyperfine coupling of the first-neighbor nitrogen atoms with high accuracy. Due to symmetry, the hyperfine tensors of the three nuclear spins can be characterized by the three hyperfine eigenstates, $A_{x}$, $A_{y}$, and $A_{z}$ as well as the azimuthal angle $\varphi$, where the latter defines the angle of the inplane $e_x$ eigenvector of the hyperfine tensor and the $x$ axis.  Only the $\varphi$ parameter is site-dependent and takes either 0, 120$^\circ$, or 240$^\circ$. In our notation, $\vartheta$ of the $e_z$ eigenvector and the electron spin quantization axis is zero. Utilizing exact diagonalization of the spin Hamiltonian  and the method detailed in the Supplementary Note~3, we fit the theoretical transition energies to the experimental results in the 187-240~mT magnetic field range. To obtain a good fit, a constant 0.947 isotropic scaling of the theoretical $^{15}$N hyperfine tensors\cite{ivady_ab_2020} was required. Thus the eigenvectors of the hyperfine tensors are $A_{x} = -61.3$~MHz,  $A_{y} = -121.6$~MHz, and $A_{z} = -63.7$~MHz. In Fig.~\ref{fig:fig_shift} and in the rest of this study, we use the fitted hyperfine tensors.

Using accurate hyperfine values in our numerical simulations, we further extend the investigated magnetic field interval to 0-500~mT, see also Supplementary Note~4. A density map of the population transfer summed up for all possible transitions is depicted in Fig.~\ref{fig:fig_shift}e. Close to the GSLAC, in the 100–150~mT region, the deviations in the transition energies $\alpha$, $\beta$, and $\gamma$ become significant and reach the 10~MHz range. Interestingly, the characteristic three-peak transition energy spectrum is also resolved on the low-field side of the GSLAC, in the 10-100~mT interval, indicating that our ENDOR protocol can be applied at these magnetic fields as well, see also Supplementary Note 6.

Note that the splitting of the hyperfine states happens not only in the $m_{\text{S}} = \pm1$ barnches but also in $m_{\text{S}} = 0$ branch due to the mixing of the states. As can be seen in Fig.~\ref{fig:fig_shift}f, the hyperfine splitting of the predominantly $m_{\text{S}} = 0$ states ranges between 2-20~MHz in the magnetic field interval of 0-500~mT. Therefore, nuclear spin state control is also possible in this RF frequency range in the $m_{\text{S}} = 0$ branch. The nuclear spin eigenstates in this branch are distinct from the $I_z$ eigenbases since the splitting merely originates from the mixing of the nuclear spin states caused by the $A_{\perp}$ hyperfine term. Consequently, these states exhibit a highly irregular magnetic field dependence in the 0-500~mT interval, see Fig.~\ref{fig:fig_shift}f.

\begin{figure*}
\includegraphics[width=1.0\textwidth]{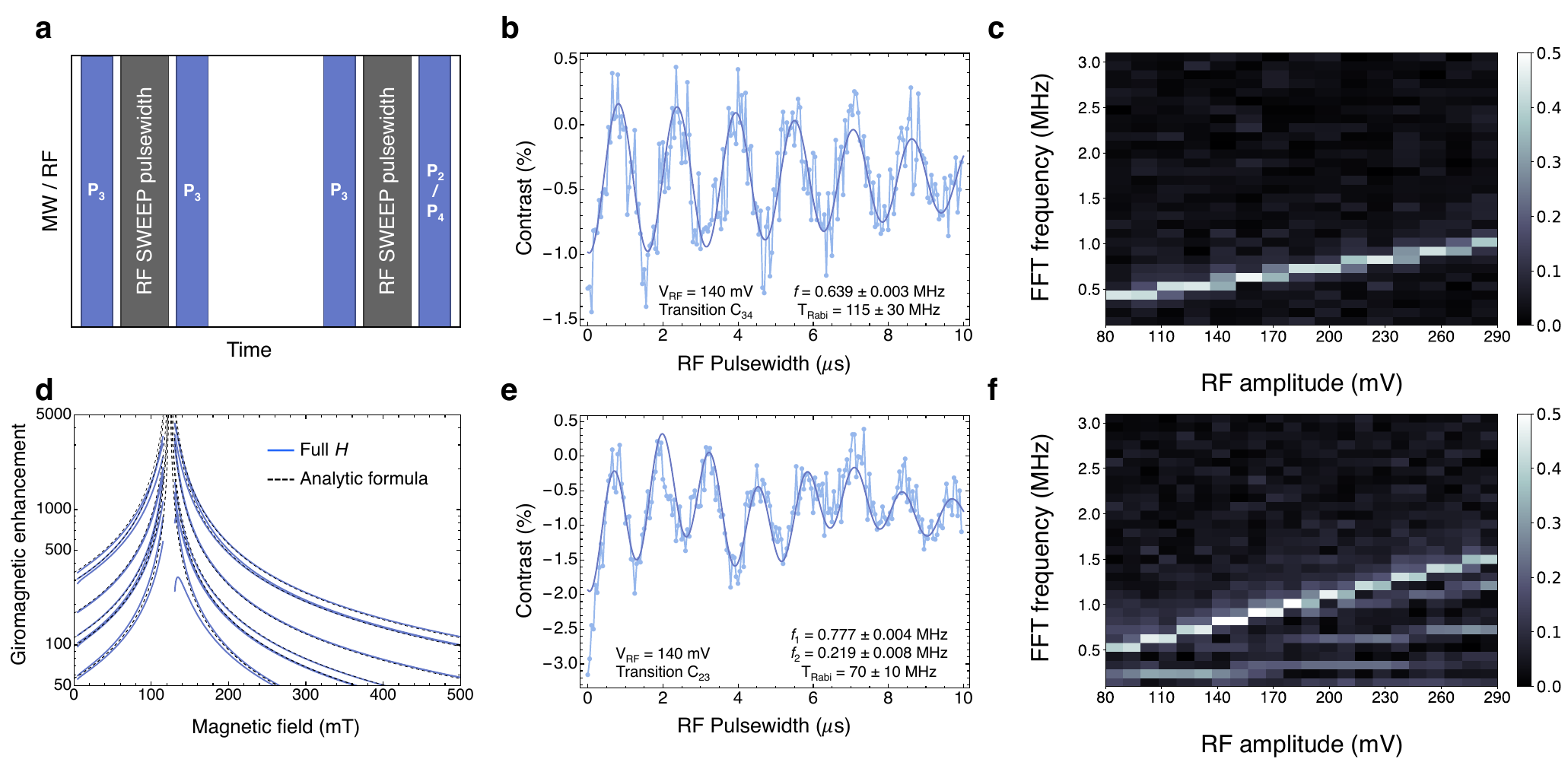}
	\caption{ \textbf{Rabi drive of nitrogen nuclear spin states.} \textbf{a} Pulse sequence used for detecting the nuclear Rabi oscillation \textbf{b} Measured Rabi oscillation between hyperfine states P$_3$ and P$_4$. \textbf{c} Fourier amplitude map (gray scale) as a function of RF drive amplitude. The frequency of the Rabi oscillation varies linearly with the RF amplitude. \textbf{d} Computed nuclear g-factor enhancement resulting from the mixing of the electron and nuclear spin states.  \textbf{e}  Rabi oscillation between hyperfine peaks P$_2$ and P$_3$, and \textbf{f} the corresponding Fourier map. For the P$_2$ $\leftrightarrow$ P$_3$ transition, the Rabi oscillation modulates due to the underlying structure of peaks P$_2$ and P$_3$ unresolved in our ENDOR spectra at the same B-field (212 mT).  }
	\label{fig:fig_rabi}  
\end{figure*}

Next, we study the Rabi drive of selected multi-nuclear spin state~\cite{gao_nuclear_2022,gong_isotope_2024} transitions $\alpha$, $\beta$, and $\gamma$. For this study, we set the external magnetic field to 212~mT and employ the pulse sequence depicted in Fig.~\ref{fig:fig_rabi}a, where the width of the RF pulse is swept. An example of the Rabi oscillation between hyperfine sublevels P$_3$ and P$_4$ (transition $\gamma$) is depicted in Fig.~\ref{fig:fig_rabi}b. The power dependence of the Rabi frequency, studied on the Fourier spectrum of the nuclear spin oscillation, is shown in Fig.~\ref{fig:fig_rabi}c. We observe a single Fourier peak whose position linearly depends on the amplitude of the RF drive in accordance with the expectations. Next, we consider Rabi oscillation between hyperfine peaks P$_2$ and P$_3$, both of which are threefold degenerate in secular approximation, but possess unresolvable fine structure at 212~mT. As seen in Fig.~\ref{fig:fig_rabi}e and f, the dominant Rabi oscillation is perturbed by the other lower frequency oscillations that we attribute to the lifted degeneracy of the three states forming peaks P$_2$ and P$_3$.

Next, we examine the dominant frequency of the Rabi oscillations. Setting the magnetic field to $B_z = 212$~mT, we drive both the electron spin state transitions between the $m_{\text{S}} = 0$ and $m_{\text{S}} = -1$ states, as well as the $\beta$ and $\gamma$ nuclear spin state transitions. Using similar MW and RF powers, the ratio of the $\Omega_e$ electron spin Rabi frequency and the $\Omega_\alpha$ nuclear spin Rabi frequency is measured to be $ \Omega_e / \Omega_\alpha \sim 56$. These values are considerably smaller than the theoretical values for an isolated $^{15}$N nuclear isotope, which is $ \Omega_e / \Omega_{^{15}\text{N}} = 6493.3$. Consequently, the Rabi oscillation of the first neighbor nuclear spin states of the V$_\text{B}^-$ center in h$^{10}$B$^{15}$N is enhanced by a factor of $\sim$116 compared to an isolated $^{15}$N nuclear spin, which is comparable to the observations of Ref.~\cite{gong_isotope_2024}. The enhancement of the nuclear g-factors is due to the non-negligible mixing of the electron and nuclear spin states in the presence of strong hyperfine interaction. Due to the large gyromagnetic ratio of the electron spin compared to the $^{15}$N nucleus, even a small mixing, see Fig.~\ref{fig:fig_shift}c, can give rise to a large increase in the strength of the effective Zeeman interaction term. To quantify this, we derive the enhancement factor in the coupled basis of the three nitrogen $^{15}$N nuclear spins, opposed to Ref.~\cite{gong_isotope_2024} where the nuclear spins are treated independently, and obtain the following formula for the largest enhancement factors for $\text{P}_2 \leftrightarrow \text{P}_3$ and $\text{P}_{1(3)} \leftrightarrow \text{P}_{2(4)}$ transitions,
\begin{equation}
    \gamma \approx  \frac{\xi}{3} \frac{g_e \mu_{\text{B}}} {g_{^{15}\text{N}} \mu_{\text{N}}}  \frac{ \left | A_{xx}^{(1)} + A_{xx}^{(2)} + A_{xx}^{(3)}  \right |  }{\left| D - g_e \mu_\text{B}B_z \right|} ,
\end{equation}
where $A_{xx}^{(i)}$ is the $xx$ element of the hyperfine tensor of nuclei $i$, $D$ is the zero field splitting, $B$ is the external magnetic field, and $\xi$ takes the value of 2 and $\sqrt{3}$ for the $+1/2 \leftrightarrow -1/2$ ($\text{P}_2 \leftrightarrow \text{P}_3$) and  the $ \pm 1/2 \leftrightarrow \pm 3/2$ ($\text{P}_1 \leftrightarrow \text{P}_2$ and $\text{P}_3 \leftrightarrow \text{P}_4$) transitions in the quartet subspace of the coupled basis, see Supplementary Note~5 for more details. The latter factor originates directly from the $\left \langle +1/2 \left | S_+^{(3/2)} \right |  -1/2 \right \rangle$ and $\left \langle \pm1/2 \left | S_+^{(3/2)} \right |  \pm3/2 \right \rangle$ matrix element of the ladder operator of the quartet spin. The ratio $\eta = 2/\sqrt{3} = 1.154$ is close to the experimental ratio of 1.216 of the dominant frequency components in Figs.~\ref{fig:fig_rabi}b and d. Altogether, we identify ten nuclear spin state transitions with $\gamma >1$, see Fig.~\ref{fig:fig_rabi}d and Supplementary Note~5. At 212~mT, the theoretical g-factor enhancement is 416, which is in the same order of magnitude as the experimental value of $\sim116$, but a factor of 3.5 larger, presumably due to the uncertainty in the MW and RF power delivered to the sample and neglecting the other nuclear spins.

From a damped exponential fit to the data in Fig.~\ref{fig:fig_rabi}b, the Rabi coherence time is $T_{Rabi}\approx12~\mathrm{\mu s}$ and trends downward with increasing RF drive amplitude. (see Supplementary Note 7).
This indicates the order of the $T_2^{*}$ for the nuclear spin states in the weak drive limit ($T_{\text{Rabi}} \approx T_{2}^{*}$ ) and is approximately two orders of magnitude longer than the T$_2$ coherence time of the electron spin.

\begin{figure*}
\includegraphics[width=1.0\textwidth]{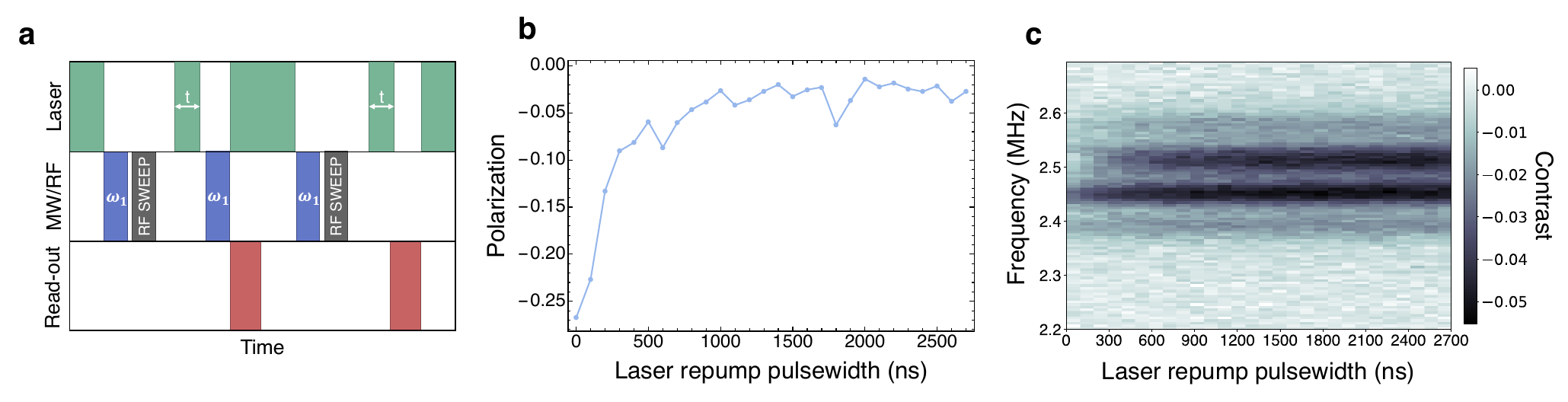}
	\caption{ \textbf{Loss of nuclear spin polarization under optical pumping.} \textbf{a} Pulse sequence used to study the repump laser duration dependence of the nuclear polarization. \textbf{b} Polarization of the $^{15}$N nuclear spin state as a function of the duration of the repump laser, see Fig.~\ref{fig:fig_shift}a. \textbf{c} ODMR map at $B = 212$~mT after employing a repump laser of varying duration.  }
	\label{fig:fig_repump}  
\end{figure*}

Finally, we study the lifetime of the nuclear spin population under continuous optical drive. The pulse sequence used in the measurements is detailed in Fig.~\ref{fig:fig_repump}a. For this measurement, we set the magnetic field to 212~mT. After populating a selected nuclear spin state with a MW and a subsequent RF pulse, we apply a laser pulse to repump the electron spin. The relative occupation of the nuclear spin states is then measured by sweeping the frequency of the microwave readout pulse for different repump pulsewidth $t$. The nuclear spin polarization is calculated from the electron spin resonance spectra by fitting a sum of four Gaussian functions. The area of the four Gaussians, $\delta_{m_I}$ are used to estimate the nuclear polarization as:  
\begin{equation}
    \mathcal{P} = \left( \sum_{m_I}m_I\delta_{m_I} \right)/\left(\frac{3}{2}\sum_{m_I}\delta_{m_I} \right) .
\end{equation}
There is a fast depolarization of the nuclear spin when the repump laser pulse is inserted (Fig. \ref{fig:fig_repump}b). The optical cycling of the  $V_B^-$ center completely thermalizes the nuclear spin within 1~$\mu$s. Apparently, this timescale is shorter than the measured order of $10~\mu \text{s}$  nuclear spin Rabi T$_2^*$ coherence time and much shorter than the expected T$_1$ time of the nuclear spins in the ground state.

Since the relaxation is a direct consequence of optical pumping, it can be assigned to the triplet optical excited $^{3}$E$^{\prime \prime}$ state and/or to the singlet $^{1}$E$^{\prime}$ shelving state. The latter state may interact with the nuclear spins with the orbital hyperfine coupling enabled by the non-zero orbital and magnetic moment of the $^{1}$E$^{\prime}$ state. On the other hand, the effective angular and, thus, the magnetic moment of the metastable state can be quenched substantially compared to a free-standing atom. For the excited state of the NV center in diamond, the effective $l$ quantum number of $L^2$ operator is measured to be $< 0.1$ \cite{rogers_time-averaging_2009}. We assume the quenching to be in a similar order for the V$_\text{B}^-$ center. Consequently, orbital hyperfine coupling is at least an order of magnitude smaller than the electron spin hyperfine coupling in the ground state. Since the stronger electron spin hyperfine coupling gives rise to slower nuclear spin relaxation in the ground state, we exclude orbital hyperfine coupling, and this way the $^{1}$E$^{\prime}$ state is the main cause of the optical pumping-induced nuclear spin relaxation. 

We thus tentatively assign the observed relaxation effects to the $^{3}$E$^{\prime \prime}$ state. While fine details of this electronic state are still not well understood, from the preliminary studies \cite{ivady_ab_2020,reimers_photoluminescence_2020}, we know that this state goes through significant Jahn-Teller distortion. Such a distortion does not imply a significant change in the hyperfine coupling\cite{gao_nuclear_2022}, however, it may lead to new electron spin relaxation pathways that can shorten the lifetime of the electron spin. It is plausible that the electron spin states' fast relaxation can affect the nuclear spin through the spin state mixing caused by the hyperfine coupling.

\section*{Discussion}

Our results demonstrate that the three nearest-neighbor $^{15}$N nuclear spins of the $V_B^-$ center in isotope-enriched hBN are not merely a decoherence source, but instead provide a controllable quantum resource. By exploiting hyperfine-mediated mixing between electron and nuclear spin states, we selectively address multi-nuclear configurations and coherently manipulate specific state pairs with well-defined Rabi oscillations. The observed enhancement of the nuclear g-factor by more than two orders of magnitude compared to isolated $^{15}$N nuclei enables rapid driving of the nuclear spin. Furthermore, the Rabi coherence times are one to two orders of magnitude longer than the electron spin $T_2$, underscoring the potential of nuclear spins mitigating decoherence in hBN-based quantum devices.

The fast control, long coherence times, and spectral selectivity of the strongly coupled nuclear spins provide a foundation for nuclear-assisted dynamics, quantum memories, and enhanced sensitivity in quantum sensing foils. The inherent 2D nature of hBN and stability of the $V_B^-$ center in few-layer samples\cite{durand_optically_2023} facilitates direct integration with van der Waals heterostructures opening new avenues for nanoscale quantum technologies for sensing\cite{daly_prospects_2025}.

\section*{Methods}
\subsection*{Sample and Device Fabrication}
Isotopically enriched $\mathrm{h^{10}B^{15}N}$ crystal flakes are grown by the atmospheric pressure high temperature (APHT) method, previously described in detail \cite{Li_JMatChemC_2020,janzen_boron_2024}. Briefly, the process starts by mixing high-purity 99.2\% enriched boron-10 with nickel and chromium with mass ratios of 3.72:48.14:48.14, respectively. The mixture is then heated at 200$\mathrm{\degree C/hour}$ under 97\% enriched nitrogen-15 and hydrogen gas at pressures of 787 and 60 torr, respectively, to 1550$\mathrm{\degree C}$, to produce a homogeneous molten solution. After 24 hours, the solution is slowly cooled at 1$\mathrm{\degree C/hour}$, to 1500$\mathrm{\degree C}$, then at 50$\mathrm{\degree C/hour}$ to 1350$\mathrm{\degree C}$, and 100$\mathrm{\degree C/hour}$ to room temperature. The $\mathrm{h^{10}B^{15}N}$ solubility is decreased as the temperature is reduced, causing crystals to precipitate on the surface of the metal. The $\mathrm{h^{10}B^{15}N}$ flakes are exfoliated from the metal with thermal release tape. The free-standing $\mathrm{h^{10}B^{15}N}$ crystalline flakes are typically greater than 20 $\mathrm{\mu m}$ thick.

Thin flakes are tape-exfoliated and dry-transferred on top of a NbTiN/Cr/Au (100/5/200 nm) coplanar waveguide (CPW) on a sapphire substrate. The NbTiN layer was included for superconducting experiments, not reported here. After transfer to the CPW sample, the hBN was implanted with 2 keV He ions at a dose of $1\times10^{14}~\mathrm{cm^{-2}}$ to induce the $V_B^-$ ensemble.

\subsection*{Experimental Setup}
The experiments were performed under ambient conditions in a home-built scanning confocal microscope. A 488 nm diode laser is focused onto the sample using a high numerical aperture objective lens (NA=0.8) and is modulated by an acousto-optic modulator. The boron vacancy photoluminescence is collected with the same objective, filtered by a 750 nm long pass filter (Thorlabs FELH0750) and coupled, via a multi-mode fiber to a single photon avalanche diode. The photon counts were recorded with time tagging electronics (Swabian Time Tagger 20). The microwave excitation was applied via the coplanar waveguide. The MW (electron) and RF (nuclear) pulses were generated by two independent channels of an arbitrary waveform generator (Keysight M8195A), amplified (Agilent 83017a for MW and Mini-Circuits TVA-R5-13 for RF), and then combined with a power combiner (Mini-Circuits ZX10R-2-183-S+). The optical and microwave pulses and PL readout were synchronised using a pulse-pattern generator (Swabian Pulse Streamer 8). A permanent magnet is mounted below the sample on an XYZ-translation stage. The magnet's XY-position is adjusted to align the magnetic field parallel to the hBN c-axis by maximizing the electron Rabi coherence time (see Supplementary Note 8). All experiments are performed under ambient conditions.

\subsection*{Numerical simulations}
For the theoretical part of this study, two main approaches were employed. The dynamics of the closed spin system were investigated through numerical simulations, with the governing equations implemented in C++. The system's time evolution was computed by solving the Liouville–von Neumann equation for the density operator, using a fourth-order Runge–Kutta integration scheme. The total simulated time was set to 600 ns, with a time step of 0.000005 ns.

The system was initialized in different quantum states to analyze nuclear spin transitions. Specifically, the density operator of the electronic spin was initially set to thermal equilibrium, while decoherence effects were neglected. Since the nuclear spin subsystem has a total of eight possible configurations, configurations with the same number of specific spin states were averaged to reduce computational complexity. The hyperfine interaction parameters used in the numerical simulations were obtained from ab initio calculations \cite{ivady_ab_2020}.

Transition matrix elements and fitted hyperfine values were computed using Mathematica. The Hamiltonian of the closed four-spin system was decomposed into two distinct components: the diagonalized Hamiltonian and the off-diagonal elements. The diagonalization was performed exactly, without any approximations. Transition matrix elements were then calculated for each pair of states.

\section*{Data availability}

The main data supporting the findings of this study are available within the paper and its Supplementary Information. Further numerical data are available from the authors upon reasonable request.

\section*{Author contributions}

I.L. conceived and performed the spin-optical experiments with C.P. The CPW samples were designed by C.P. and fabricated by B.L. hBN flakes were synthesized by T.P. and J.E. A.T. and V.I. carried out the theoretical study and the computer simulations. The paper was written by V.I., A.T., and I.L. with inputs from all authors. The work was supervised by I.L. and V.I.

\section*{Competing interests} 

The authors declare no competing interests.

\section*{Acknowledgments}

 We thank Iman Esmaeil Zadeh for useful discussions and guidance in device fabrication. This research was supported by the National Research, Development, and Innovation Office of Hungary within the Quantum Information National Laboratory of Hungary (Grant No. 2022-2.1.1-NL-2022-00004) and grant FK 145395. We acknowledge the support of the European Union under Horizon Europe for the QRC-4-ESP project (Grant Agreement 101129663) and the QUEST project (Grant Agreement 101156088). The computations were enabled by resources provided by the National Academic Infrastructure for Supercomputing in Sweden (NAISS) and the Swedish National Infrastructure for Computing (SNIC) at NSC, partially funded by the Swedish Research Council through grant agreement no. 2022-06725 and no. 2018-05973. Support for the hexagonal boron nitride crystal growth was provided by the National Science Foundation, award number 2413808.

\clearpage

\printbibliography

\end{document}


\pagestyle{fancy}
\rhead{\includegraphics[width=2.5cm]{figs/VCH-logo.png}}

\title{Supplementary Information for the paper entitled \\ \vspace{0.5 cm} Coherent control of nitrogen nuclear spins via the V$_B^-$-center in hexagonal boron nitride}

\maketitle


\author{Adalbert Tibiássy, Charlie J. Patrickson, Thomas Poirier, James H. Edgar, Bruno Lopez-Rodriguez, Iman Esmaeil Zadeh, Viktor Ivády, Isaac J. Luxmoore}

\begin{affiliations}
Adalbert Tibiássy,  Viktor Ivády \\
\emph{Department of Physics of Complex Systems, E\"otv\"os Lor\'{a}nd University,\\
Egyetem t\'{e}r 1-3, H-1053 Budapest, Hungary, and } \\
\emph{MTA–ELTE Lend\"{u}let "Momentum" NewQubit Research Group,\\
P\'{a}zm\'{a}ny P\'{e}ter, S\'{e}t\'{a}ny 1/A, 1117 Budapest, Hungary}\\

\vspace{0.4 cm}

Charlie J. Patrickson and Isaac J. Luxmoore \\
\emph{Department of Engineering, University of Exeter, EX4 4QF, United Kingdom}\\

\vspace{0.4 cm}

Thomas Poirier and James H. Edgar \\
\emph{Tim Taylor Department of Chemical Engineering, Kansas State University, Manhattan, Kansas 66506, USA}\\

\vspace{0.4 cm}

Bruno Lopez-Rodriguez and Iman Esmaeil Zadeh\\
\emph{Department of Imaging Physics (ImPhys), Faculty of Applied Sciences, Delft University of Technology, 2628 CJ Delft, The Netherlands}\\

\vspace{0.4 cm}

Email Address: ivady.viktor@ttk.elte.hu and i.j.luxmoore@exeter.ac.uk

\end{affiliations}


\keywords{quantum sensing, VB center in hBN, decoherence, coherence protection}

\justifying

\newpage

\section{Supplementary Note 1 - Spin Hamiltonian}

A total of four spins are involved in the model examined in this paper: the triplet electron spin of the V$_\text{B}^-$ center and the three nearest neighbour spin-1/2 $^{15}$N nuclear spins, which are strongly coupled to the electron spin. The Hilbert space on which the spin operators act has a dimension of $3 \cdot 2^3 = 24$.
The complete \textit{time-dependent} spin Hamiltonian operator of the system can be constructed using the following interaction terms: 
\begin{equation} 
        \hat{H}(t) = \hat{H}_{\text{ZFS}}^{(e^-)} + \hat{H}_{\text{Zeeman}}^{(e^-)}(t) +\sum_i^3 \left( \hat{H}_{\text{Zeeman}}^{(\text{nuc}, i)}(t) + \hat{H}_{\text{hyperfine}}^{(e^- - \text{nuc},i)}\right), 
\end{equation}
where $\hat{H}_{\text{ZFS}}^{(e^-)} $ is the Hamiltonian operator corresponding to the zero-field splitting, $\hat{H}_{\text{Zeeman}}^{(e^-)}$(t) and $\hat{H}_{\text{Zeeman}}^{(\text{nuc},i)}$(t) describe the Zeeman interaction of the electron and the \textit{i}th nuclear spin, respectively. The operators $\hat{H}_{\text{HFS}}^{(e^- - \text{nuc},i)}$ describe the hyperfine interaction between the vacancy center and the \textit{i}th nuclear spin. The individual terms of the complete Hamiltonian are detailed below.

\subsubsection*{Zero-field splitting}
Zero-field splitting appears when the system contains two or more unpaired electrons, meaning that $S \geq 1$. The Hamiltonian operator for zero-field splitting can be written as
\begin{subequations} 
    \begin{equation}
        \hat{H}_{\text{ZFS}} = \mathbf{S} \hat{D} \mathbf{S} = D_{\text{x}} S_{\text{x}}^2 + D_{\text{y}} S_{\text{y}}^2 + D_{\text{z}} S_z^2,
    \end{equation}
where $D_{\text{x}}$, $D_{\text{y}}$, and $D_{\text{z}}$ are the eigenvalues of the diagonalized $\hat{D}$ tensor.
Introducing a new notation for the constants $D = 3 D_{\text{x}}/2$ and $E = (D_{\text{x}} - D_{\text{y}})/2$, the zero-field splitting Hamiltonian becomes 
    \begin{equation}\label{eq:hamilton_E_parameter}
        \hat{H}_{\text{ZFS}} = D \left( S_{\text{z}}^2 - \frac{S(S+1)}{3} \mathbb{1}_3 \right) + E \left( S_{\text{x}}^2 - S_{\text{y}}^2 \right). 
    \end{equation}
\end{subequations}
For the V$_\text{B}^-$ center in boron nitride, $D = 3479$ MHz and $E = 50$ MHz. The $E$ parameter  does not appear if the system is completely symmetric; however, this term is not zero in experiments.

\subsubsection*{Zeeman terms}
In our case, the external magnetic field has two components:
a) a static $B_0$ term aligned parallel to the quantization axis, and
b) a time-dependent component $B(t)$ perpendicular to $c$.
By convention, the quantization axis is the $z$-axis, which is parallel to the symmetry axis. Thus, the Zeeman interaction term for the electron spin can be written as
\begin{equation} \label{eq:zeeman_1}
    \hat{H}_{\text{Zeeman}}^{(e^-)}(t) = \gamma_e B_0  S_{\text{z}}+ \gamma_e B_{\text{A}} \cos(\omega t) S_{\text{x}} ,
\end{equation}
where $g_e$ is the electron Landé g-factor, $\mu_B$ is the Bohr magneton, and $\omega$ and $B_{\text{A}}$ are the frequency and amplitude of the oscillating magnetic field.

The Zeeman terms acting on the nitrogen nuclear spins are very similar to the Zeeman Hamiltonian term of the electron. The only difference is that the multiplicative factor is given by the nuclear $\mu_N$ nuclear magneton and the Landé g-factor of the $^{15}$N isotopes, denoted as $g_{^{15}\text{N}}$. The Hamiltonian operator of the nuclear Zeeman interaction for a given nuclear spin is
\begin{subequations} 
    \begin{equation}\label{eq:zeeman_4} 
        \hat{H}_{\text{Zeeman}}^{(\text{nuc},1)}(t) = \gamma_{^{15}N} B_0 I_{\text{z}} + \gamma_{^{15}N}  I_{\text{x}} B_{\text{A}} \cos(\omega t), 
    \end{equation}
\end{subequations}
where $\hat{I}_{\text{z}}$ and $\hat{I}_{\text{x}}$ are the $z$ and $x$ operators of the nuclear spin. 
The value of the nuclear magneton is $7.6228 \cdot 10^{-4}$ Am$^2$, and the gyromagnetic factor of the $^{15}$N is $g_{^{15}N} = -0.56638$.

The time-dependent Zeeman Hamiltonian operator provides a semiclassical approximation of the MW/RF driven V$_{\text{B}}^-$ center, as the quantization of the electromagnetic field is not considered. This is referred to as the "strong-field" approximation because it assumes that the absorption and emission of a photon by the V$_\text{B}^-$ center does not significantly alter the properties of the field itself.

\subsubsection*{Hyperfine splitting}
The hyperfine splitting originates from the magnetic dipole interaction of the electron and the nuclear spin, which can be expressed as
\begin{equation}
    \hat{H}_{\text{dip}} = \gamma_e \gamma_{^{15}N} \dfrac{1}{|\mathbf{r}-\mathbf{R}|^3}\left( \dfrac{\mathbf{S}(\mathbf{r}-\mathbf{R}) \cdot \mathbf{I}(\mathbf{r}-\mathbf{R})}{|\mathbf{r}-\mathbf{R}|} - \mathbf{S} \mathbf{I}\right), 
\end{equation}
where $\mathbf{r}$ and $\mathbf{R}$ are the position vectors of the electron and the nuclear spin, while $\mathbf{S}$ and $\mathbf{I}$ are the electron and nuclear spin operator vectors, respectively. After integrating over the spatial variables, the operator of the hyperfine interaction can be written in the form
\begin{subequations}
    \begin{equation}\label{eq:A_tenzor}
        H_{\text{hyperfine}}^{(\text{nuc},i)}=
        \langle \mathbf{r} | \hat{H}_{\text{dip}} |\mathbf{r} \rangle = \dots = \mathbf{S} \hat{A}_i \mathbf{I}_i = 
    \end{equation}
    \begin{equation}\label{eq:hyperfine_interaction}
         = \begin{pmatrix}
            S_x &S_y & S_z
        \end{pmatrix} 
        \begin{pmatrix}
            A_{xx,i} && A_{xy,i} && A_{xz,i} \\  A_{xy,i} && A_{yy,i} && A_{yz,i} \\  A_{xz,i}&&  A_{yz,i}&& A_{zz,i} 
        \end{pmatrix} \begin{pmatrix}
            I_{x,i}  \\ I_{y,i}  \\  I_{z,i} 
        \end{pmatrix}.
    \end{equation}
\end{subequations}
The matrix $\hat{A}_i$ appearing in equation (\ref{eq:A_tenzor}) is called the hyperfine tensor, which can be obtained from first-principles calculations\cite{takacs_accurate_2024}.

The parameters of the hyperfine tensors used in the calculations are taken from Ref.~\cite{ivady_ab_2020}. Since this source provides the hyperfine tensors for the $^{14}$N nitrogen isotop, the values must be rescaled by the ratio of gyromagnetic factors of $^{15}$N and $^{14}$N to perform calculations for vacancy centers surrounded by $^{15}$N nuclei.

\section{Supplementary Note 2  - Perturbation Theory and Analytical Considerations on State Mixing}

As can be seen in Fig.~1d of the main text, the energy gap between the different nuclear spin states is not equal in the $m_S=-1$ subspace. The gaps also change with the increase of the magnetic field, see Fig.~2 of the main text, which can be explained with perturbation theory considerations. In our case, the static Hamiltonian can be written as a sum of a completely diagonal operator $\hat{H}_0$ and a completely off-diagonal perturbation operator  $\hat{H}_p$ which can be expressed as
\begin{equation}
    \hat{H}_0  = D \left( S_{\text{z}}^2 - \frac{S(S+1)}{3} \mathbb{1}_3 \right) +  \gamma_e B_0  S_{\text{z}} + \sum_{i = 1}^3  \gamma_{^{15}N} B_0 I_{\text{z}}  + S_z \sum_{i = 1}^3  A_{zz,i} I_{z,i}.
\end{equation}
and
\begin{equation}
    \hat{H}_p  = E \left( S_x^2 + S_y^2 \right) +  S_x \sum_{i = 1}^3 \left[ A_{xy,i} I_{y,i} + A_{xx,i} I_{x,i} \right] + S_y \sum_{i = 1}^3 \left[ A_{xy,i} I_{x,i} + A_{yy,i} I_{y,i} \right].
\end{equation}
At high external $B$ fields, the mixing effect of the zero-field splitting interaction can be neglected; thus, the $E$ parameter is considered to be zero here. 

The energy correction to the diagonal terms due to the mostly suppressed mixing terms can be expressed as
\begin{equation}\label{eq:perturbation}
    E_n = E_n^{(0)} + \underbrace{\mel{n}{\hat{H}_p}{n}}_{\text{1st order PT}} +  \underbrace{ \sum_{k \neq n} \dfrac{ \left| \mel{n}{\hat{H}_p}{k} \right|^2 }{E_n^{(0)} - E_k^{(0)}}}_{\text{2nd order PT}} + \dots,
\end{equation}
where the $n$ denotes the $n$th eigenstate of the diagonal $H_0$ matrix in the $z$ basis. The 1st order energy correction is zero because the $\hat{H}_p$ contains only off-diagonal terms. The second-order correction gives a non-zero correction, which, after some algebra, can be written as
\begin{equation*}
    E_{\ket{-1;\uparrow;\uparrow;\uparrow}}^{(2)} = \dfrac{3}{8} \cdot \dfrac{ \left( A_{xx} - A_{yy} \right)^2 + 4 A_{xy}^2 }{ D - g_e \mu_B B + g_N \mu_N B - \frac{3}{2} A_{zz} }
\end{equation*}
\begin{equation*}
E_{\ket{-1;\downarrow;\downarrow;\downarrow}}^{(2)} = \dfrac{3}{8} \cdot \dfrac{ \left( A_{xx} - A_{yy} \right)^2 + 4 A_{xy}^2 }{ D - g_e \mu_B B + g_N \mu_N B+ \frac{3}{2} A_{zz} }
\end{equation*}
\begin{equation*}
    E_{\ket{-1;\downarrow;\uparrow;\uparrow}}^{(2)} = \dfrac{1}{8} \cdot \dfrac{ \left( A_{xx} + A_{yy} \right)^2}{ D - (g_e \mu_B  - g_N \mu_N) B - \frac{1}{2} A_{zz} } + 
        \dfrac{2}{8} \cdot \dfrac{ \left( A_{xx} - A_{yy} \right)^2 + 4 A_{xy}^2 }{ D - g_e \mu_B B + g_N \mu_N B - \frac{1}{2} A_{zz} }
\end{equation*}
\begin{equation*}
    E_{\ket{-1;\downarrow;\downarrow;\uparrow}}^{(2)} = \dfrac{2}{8} \cdot \dfrac{ \left( A_{xx} + A_{yy} \right)^2}{ D - (g_e \mu_B + g_N \mu_N) B - \frac{1}{2} A_{zz} } + 
        \dfrac{1}{8} \cdot \dfrac{ \left( A_{xx} - A_{yy} \right)^2 + 4 A_{xy}^2 }{ D - (g_e \mu_B - g_N \mu_N) B - \frac{1}{2} A_{zz} }
\end{equation*}
As can be seen, all the corrections are state-dependent and magnetic field-dependent, which qualitatively explains the experimental observations, i.e.\  the non-equivalence of the $\alpha$, $\beta$, and $\gamma$ transitions and their magnetic field dependence. On the other hand, we conclude from the comparison with the experiments that second-order perturbation theory is not sufficient to obtain a qualitative explanation, and fourth-order energy corrections are needed. Due to the complexity of the 4th order, it is not advantageous to provide a closed form in this order. Exact diagonalization provides quantitatively good results for the transition energies, see the next section.

By ignoring the hyperfine structure of the $m_{\text{S}} = \pm1$ subspace, a formula for the mixing of the electron spin states can be obtained by second-order consideration. The dashed curve plotted in Fig.~2c captures the leading order of the state mixing is obtained as
\begin{equation}
   p_{m_S=\pm1 \text{ in } m_S=0 \text{ branch} } = \frac{1}{2} \frac{ \left( A_{xx} + A_{yy} \right)^2 + 4A_{xy}^2}{\left(  D + \gamma_e B \right)^2} + \frac{1}{2} \frac{ \left( A_{xx} + A_{yy} \right)^2 + 4A_{xy}^2}{\left(  D - \gamma_e B \right)^2}  .
\end{equation}


\section{Supplementary Note 3 - Fitting the Hyperfine Tensor}

As a starting point for numerically simulating the measured curves, we utilized ab initio hyperfine tensors published in Ref.~\cite{ivady_ab_2020}. At first, these values are scaled by the ratio of the nuclear g-factors $g_{^{15}\text{N}} / g_{^{14}\text{N}}$ for our h$^{10}$B$^{15}$N system. However, subsequent analysis revealed that these data did not provide quantitatively adequate agreement with experimental observations, necessitating a fitting procedure to the measured data. To accomplish this, we computed the matrix elements of the perpendicular to $c$ Zeeman Hamiltonian for the $C_{22}$ and $C_{33}$ transitions using the model Hamiltonian as
\begin{equation}
     \mathcal{M}_{fi}  =  \left \langle \psi_f   \left | \gamma_e  S_{\text{x}} + \sum_{i = 1}^3  \gamma_{^{15}N} I_{\text{x}} \right | \psi_i \right \rangle.
\end{equation}

\begin{figure}
    \centering
    \includegraphics[width=\linewidth]{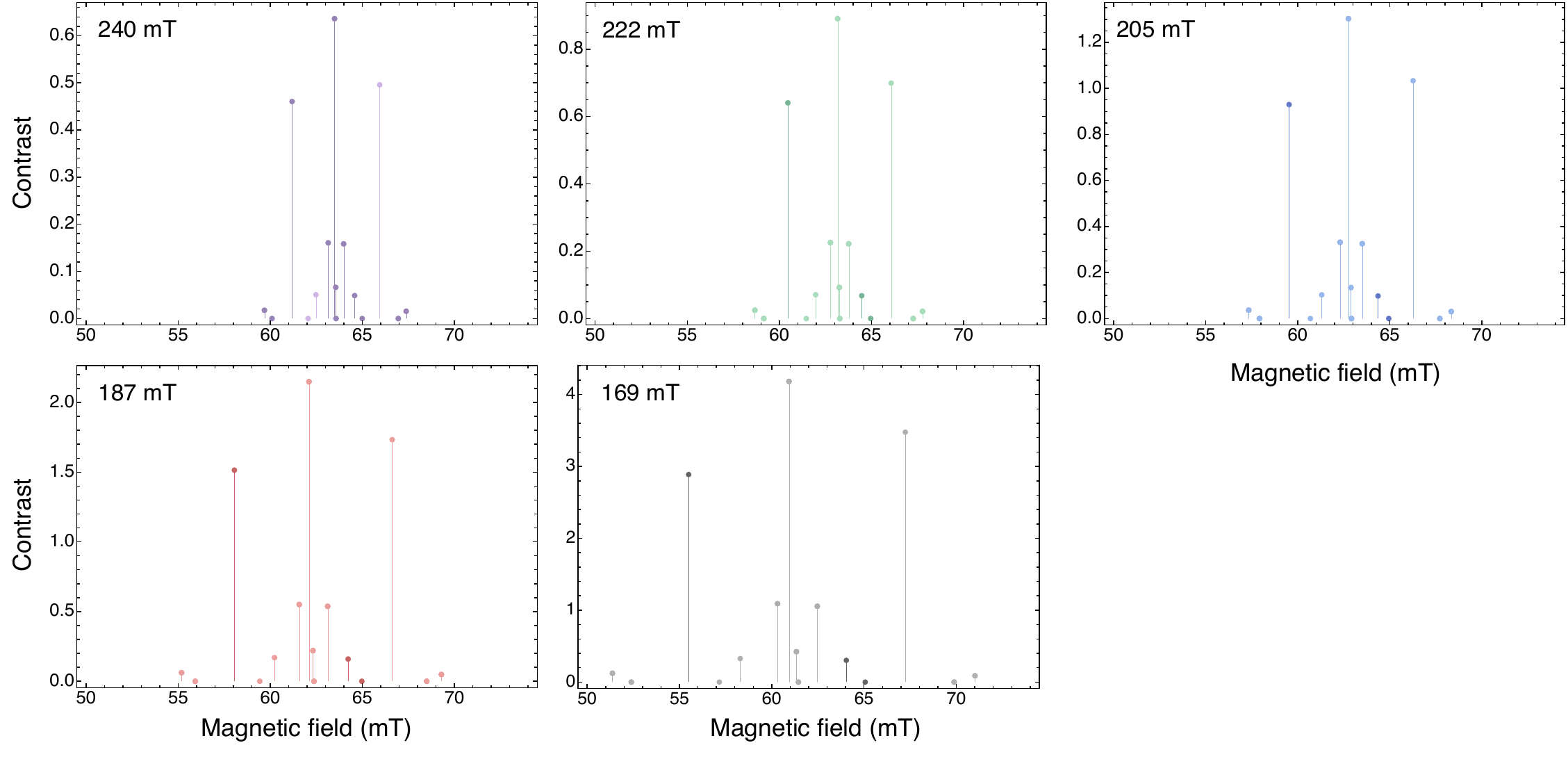}
    \caption{The calculated transition matrix elements plotted against the transition energies at different magnetic field values. }
    \label{fig:matrixelements_peaks}
\end{figure}

\begin{figure}
    \centering
    \includegraphics[width=\linewidth]{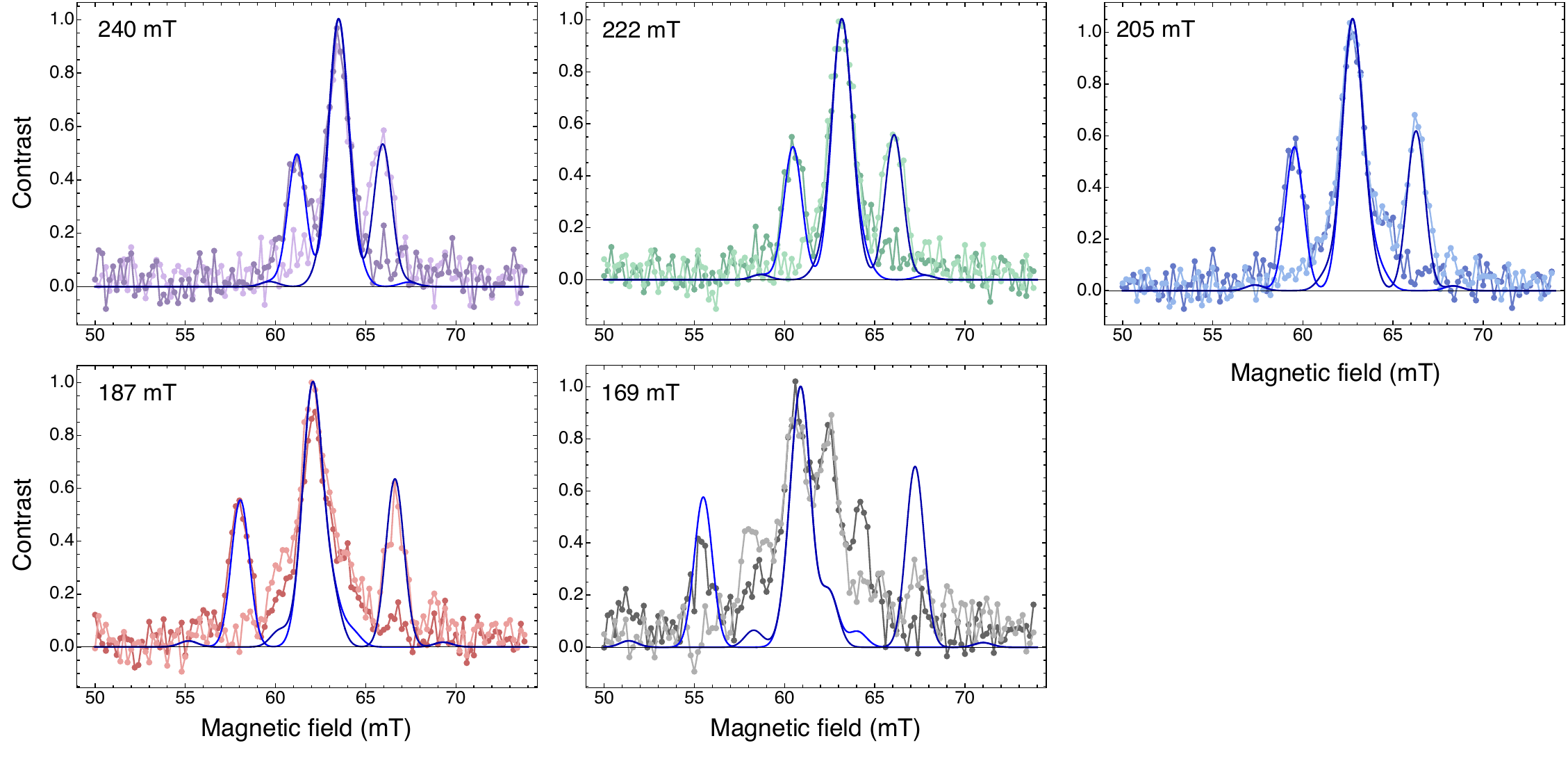}
    \caption{ Experimental ENDOR spectra compared with numerical simulations. The theoretical curves are obtained by convolving peaks plotted in Fig.~S\ref{fig:matrixelements_peaks} with a Gaussian.}
    \label{fig:fitting}
\end{figure}

The resulting transition matrix elements are depicted against the $\varepsilon_{fi}$ transition energy for various magnetic field values in Fig.~S\ref{fig:matrixelements_peaks}. To facilitate comparison with experimental data, we applied Gaussian broadening to each matrix element. The Gaussian distribution was characterized by a standard deviation of $\sigma = 0.5$ MHz, corresponding to a full width at half maximum (FWHM) of 1.1775 MHz. The convolution of these broadened peaks yields a spectrum that closely resembles experimental measurements, see Fig.~S\ref{fig:fitting}.

To achieve a good agreement between the theory and experiment, as shown in Fig.~S\ref{fig:fitting}, we scaled the \emph{ab initio} hyperfine tensors as
\begin{equation}
    \hat{A}_1 = p_\text{fitting} \cdot 
    \dfrac{g_{^{15}\mathrm{N}}}{g_{^{14}\mathrm{N}}} \cdot 
    \begin{pmatrix}
        80.202  & -19.687 & 0 \\
        -19.687 & 57.479  & 0 \\
        0       & 0       & 47.935
    \end{pmatrix}
\end{equation}
\begin{equation*}
    \hat{A}_2 = p_\text{fitting} \cdot 
    \dfrac{g_{^{15}\mathrm{N}}}{g_{^{14}\mathrm{N}}} \cdot 
    \begin{pmatrix}
        46.110  & 0 & 0 \\
        0 & 91.571  & 0 \\
        0       & 0       & 47.935
    \end{pmatrix}
\end{equation*}
\begin{equation*}
    \hat{A}_3 = p_\text{fitting} \cdot 
    \dfrac{g_{^{15}\mathrm{N}}}{g_{^{14}\mathrm{N}}} \cdot 
    \begin{pmatrix}
        80.202  & 19.687 & 0 \\
        19.687 & 57.479  & 0 \\
        0       & 0       & 47.935
    \end{pmatrix}
\end{equation*}
where the $p_\text{fitting}$ is the fitting parameter used for the correction of the hyperfine tensor elements. With a reasonably scaling factor, the theoretical curves almost perfectly fit the experiments. The value of the  scaling factor is found to be $p_\text{fitting} = 0.947$. 

As shown in Fig \ref{fig:fitting}, the theoretical curve at the lowest magnetic field does not match as well as for higher magnetic field values. This can be explained by the limitation of the model since not all nuclear spins are included in the Hamiltonian, which might have a stronger influence on the curves near the GSLAC.





\section{Supplementary Note 4 - Numerical simulations on transition strengths}

\begin{figure}
    \centering
    \includegraphics[width=\linewidth]{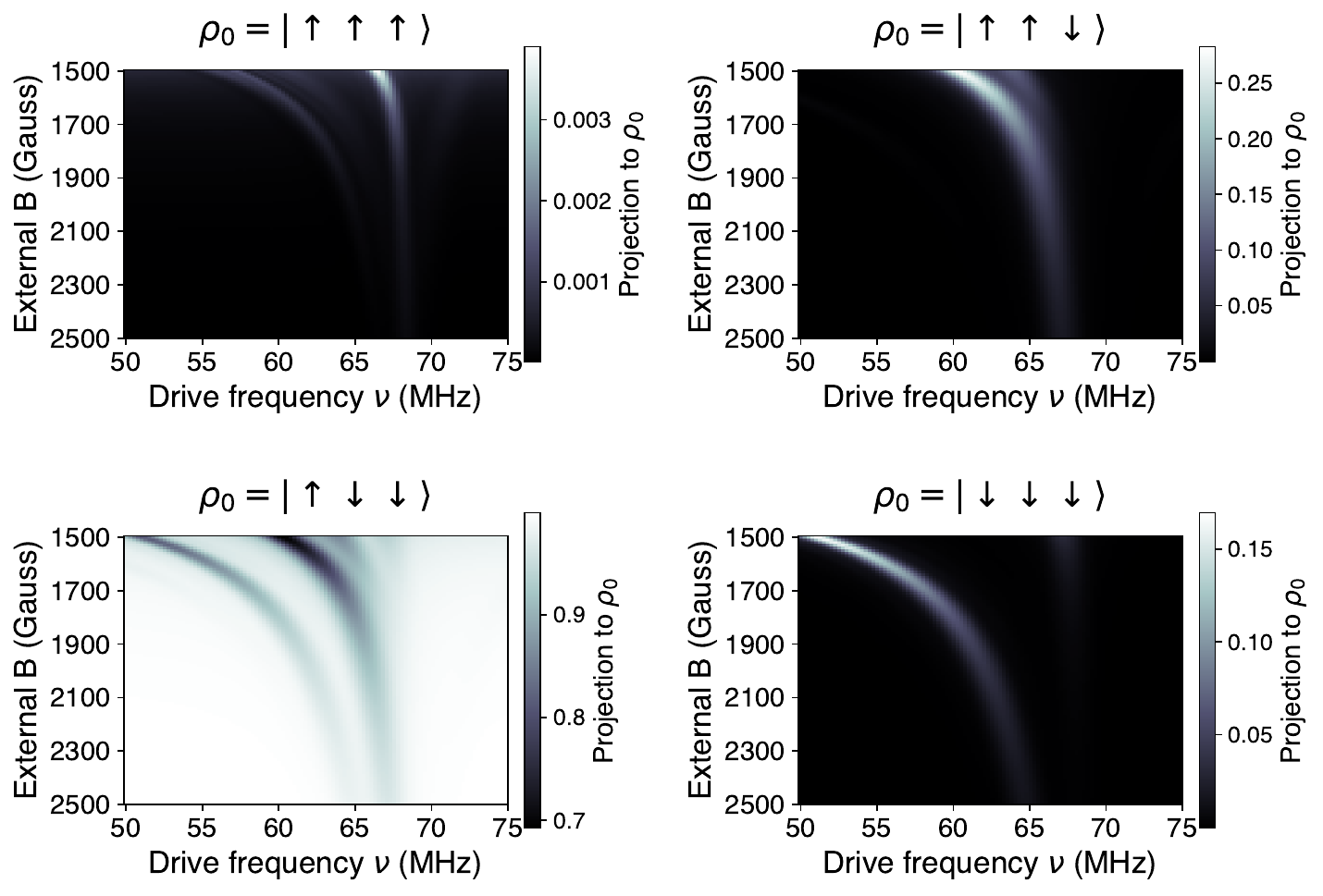}
    \caption{
    The system was specifically initialized in the $\ket{-1, \uparrow \downarrow \downarrow}$ state, corresponding to the P2 configuration. Consequently, the third figure exhibits higher intensity compared to the remaining three, as this figure displays the projection calculated with respect to the initial state $\ket{-1, \uparrow \downarrow \downarrow}$. The other three figures show projections onto the alternative states. Notably, a projection onto the $\ket{-1, \uparrow \uparrow \uparrow}$ state is also observable, despite this configuration being separated from the initial state by more than two spin flip transitions.}
    \label{fig:init_to_udd}
\end{figure}

In this section, we provide density plots of the transition strength obtained by time-dependent numerical simulation of a closed-four-spin system.  The simulations started from the definite $\varrho_0$ state, and the probability of transition is tested by projecting the time-dependent density operators either onto the initial density or onto a projector operator $P$ and integrating over time as
\begin{equation}
    \Xi = \int_0^T \text{Tr} \left( P \varrho  (t) \right) dt.
\end{equation}
Considering the different initial spin configurations possible, there are altogether four relevant ones.

\begin{figure}
    \centering
    \includegraphics[width=\linewidth]{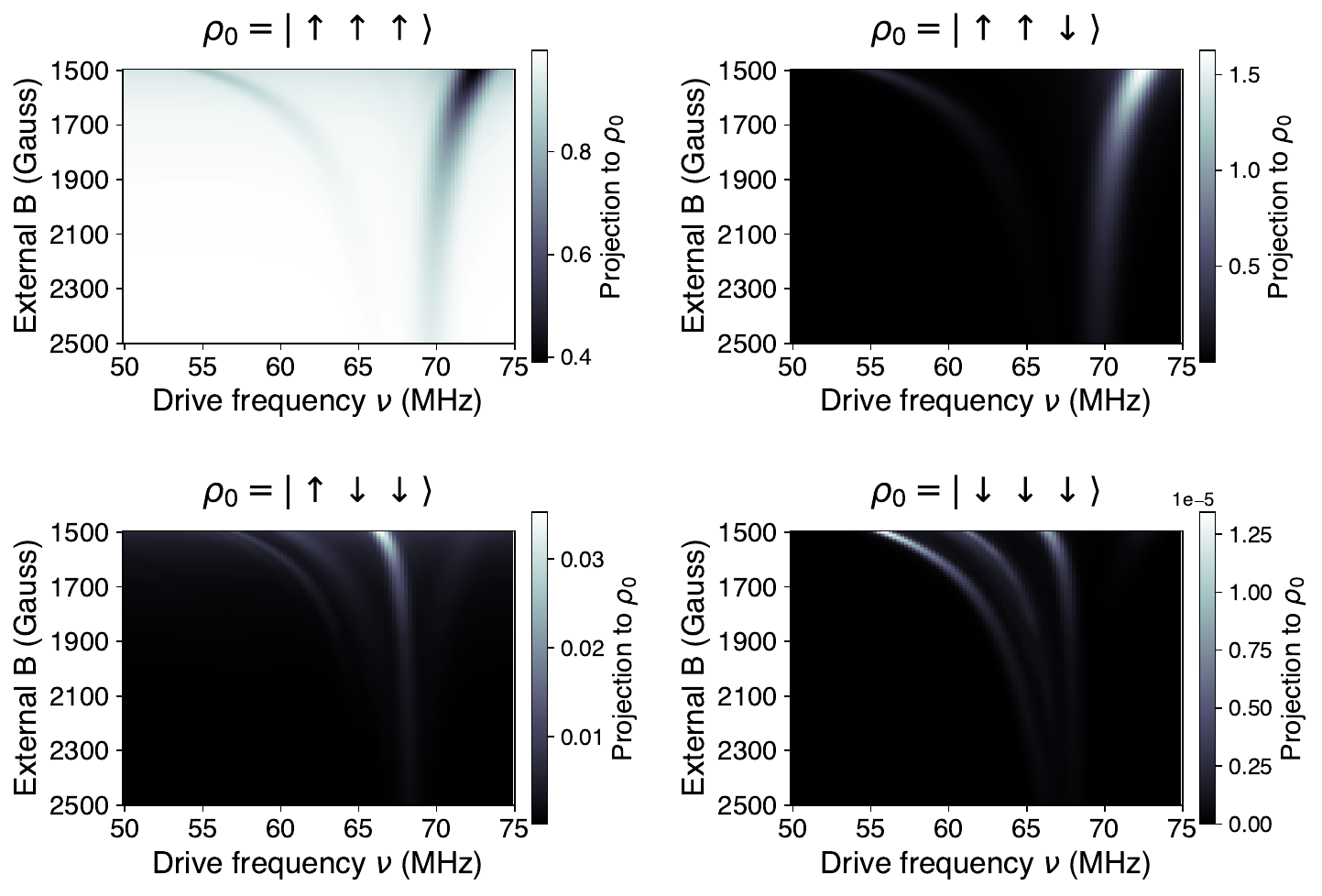}
    \caption{The same process as shown in Fig.~S\ref{fig:init_to_udd} was employed, with the  difference that the system was initialized in the $\ket{-1,\downarrow \downarrow \downarrow}$ stat, and each subfigure shows the projection to the given $\varrho_0$.}
    \label{fig:init_to_ddd}
\end{figure}

Figs.~\ref{fig:init_to_udd}-\ref{fig:init_to_all_long} depict and discuss the results of the numerical simulations.

\begin{figure}
    \centering
    \includegraphics[width=\linewidth]{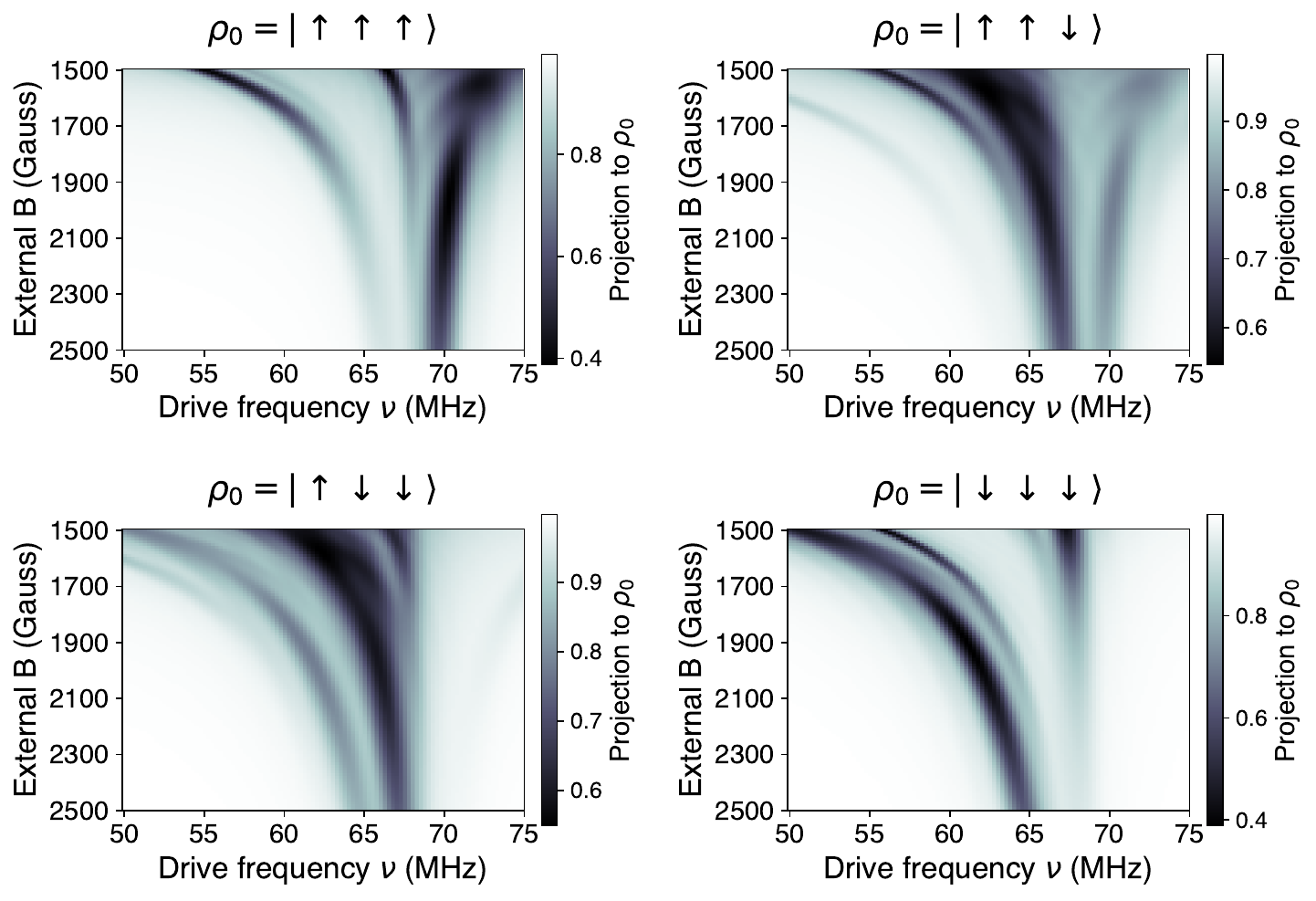}
    \caption{In this figure, each subfigure corresponds to simulation results where the system is initialized in $\varrho_0$, and projected onto $\varrho_0$. Transitions appear as black bands.}
\end{figure}

\begin{figure}
    \centering
    \includegraphics[width=\linewidth]{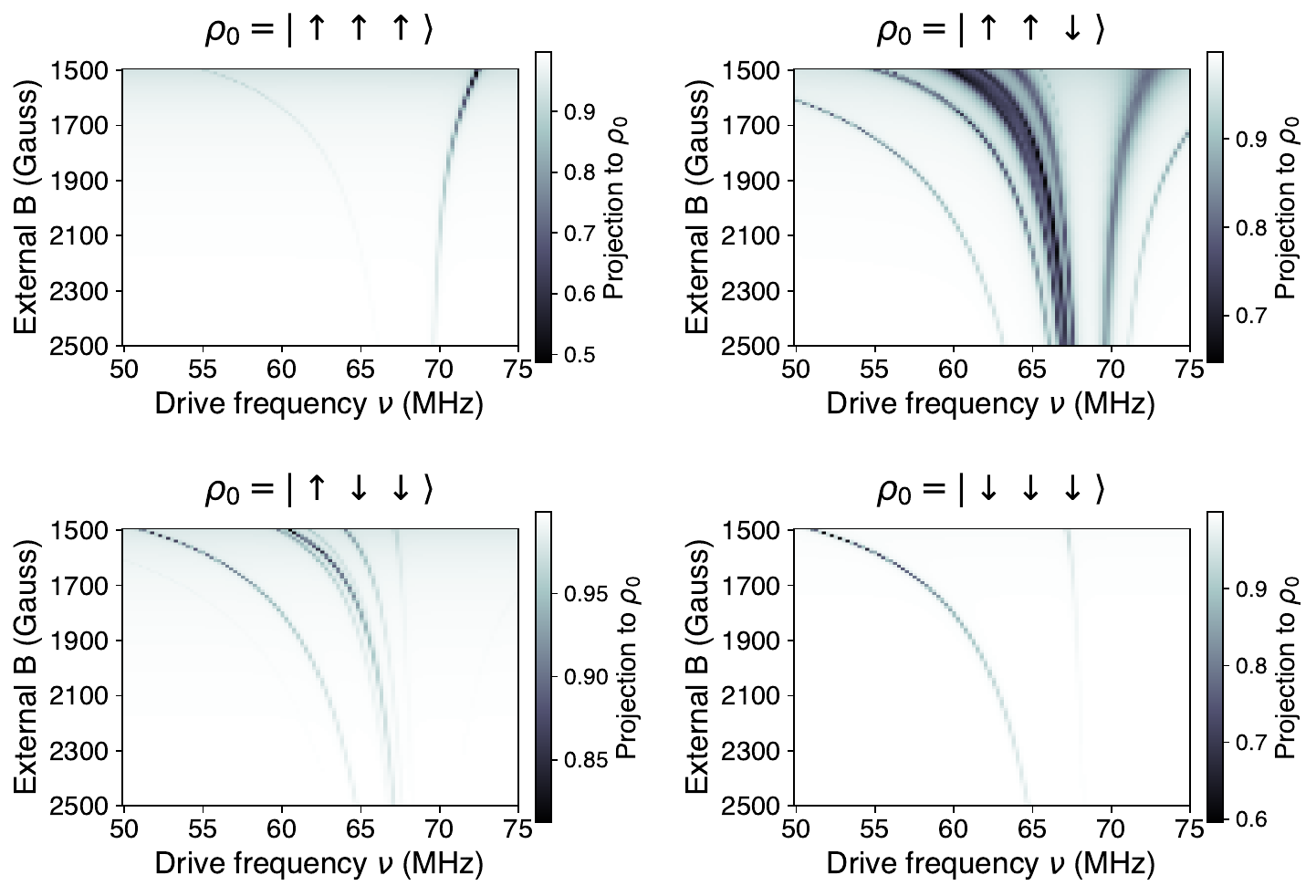}
    \caption{The system was time-evolved over an extended period of 600 ns, in contrast to the shorter durations employed in previous simulations. This extended simulation time results in enhanced separation and clarity of the transition lines, as the temporal averaging of the system's occupation in different states increases proportionally with the simulation duration.}
    \label{fig:init_to_all_long}
\end{figure}

\section{Supplementary Note 5 -  g-factor Enhancement}

We show in the main text that the mixing of electron and nuclear spin states plays an important role for the V$_\text{B}^-$ center in h$^{11}$B$^{15}$N. Obvious consequences of this mixing are the shift and splitting of energy levels, as well as the modulation of magnetic interaction terms. This effect is most striking for the nuclear spin states, where even a small degree of mixing gives rise to significant coupling with external magnetic fields, due to the orders-of-magnitude difference between the electron and nuclear magnetic moments. The enhanced magnetic coupling of nuclear spin is often described by its effective nuclear $g$-factor $g_{\text{eff}}$, and the nuclear $g$-factor enhancement $\gamma = g_{\text{eff}} / g_{\text{N}}$. In experiments, however, often only the ratio $\varepsilon$ of the electron $g$-factor and the effective nuclear $g$-factor can be measured, owing to the unknown MW/RF power delivered to the sample. In this supplementary section, we derive these quantities for the states of the first three neighboring $^{15}$N nuclei of the V$_\text{B}^-$ center in h$^{11}$B$^{15}$N.

In our setup, a coplanar waveguide delivers the MW and RF drive to the sample. In such circumstances, the magnetic field can be approximated as $\vec{B}(t) \simeq \left\lbrace B_x(t), 0, B_z \right\rbrace$, where $B_x(t) = B_{\text{drive}} \cos(\omega t)$. The Rabi frequency for a transition between $\left| i \right\rangle$ and $\left| f \right\rangle$ is given by
\begin{equation}
    \Omega = \frac{1}{\hbar} \left| \left\langle f \middle| H_{\text{Zeeman}} \middle| i \right\rangle \right|,
\end{equation}
where
\begin{equation}
    H_{\text{Zee}} = g_e \mu_{\text{B}} B_x \hat{S}_x + g_{^{15}\text{N}} \mu_{\text{N}} B_x \sum_{j=1}^3 \hat{I}_x^{(j)}
\end{equation}
is the interaction Hamiltonian describing the coupling of the four-spin system to the time-dependent magnetic field.

For a pure electronic and nuclear spin state, the Rabi frequency is 
\begin{equation}
    \Omega_{\text{el}} = \frac{g_e \mu_{\text{B}} B_{\text{drive}}}{\sqrt{2}}
\end{equation}
and
\begin{equation}
    \Omega_{^{15}\text{N}} = \frac{g_{^{15}\text{N}} \mu_{\text{N}} B_{\text{drive}}}{2}
\end{equation}
respectively, where $\sqrt{2}$ and the 2 factors in the denominator originate from the matrix elements of $\hat{S}_x$ and $\hat{I}_x$ respectively. 

To obtain the g-factor enhancement and the ratio of the electron and nuclear Rabi frequencies in our four-spin system, multiple pairs of states can be used for the initial $\left| i \right\rangle$ and the final $\left| f \right\rangle$ nuclear spin states. From now on, we will exclusively consider the $m_{\text{S}} = -1$ electron spin branch, which altogether includes $2^3 = 8$ nuclear spin states. As an important difference from the derivation presented in the literature in Ref.~\cite{gong_isotope_2024}, here we use the coupled $\left| j,m_j \right\rangle$ basis of the three nuclear spins, rather than considering them as three independent nuclear spins. This has important consequences for the final formulas and gives results that fit our \vi{experiments better}. The choice of basis is motivated by the indirect coupling of the nuclear spins through the hyperfine interaction with the electron spins.

The eight-dimensional Hilbert space of the three $^{15}$N nitrogen divides into two doublet subspaces, D$_1$ and D$_2$, and a quartet subspace Q in the coupled $\left| j,m_j \right\rangle$ basis. The subspaces are spanned by the  states listed in Table~\ref{sitab:states} expressed in $\sum \hat{I}_z$ basis.

\begin{table}[!h]
    \centering
    \caption{List of doublet and quartet eigenstates expressed in $\hat{I}_z$ basis. The coupled basis states are used to express the matrix elements of the Zeeman and hyperfine terms in the study of the g-factor enhancement factor calculations.}
    \label{sitab:states}
    \begin{tabular}{c |c|c}
    \hline
    subspace & basis state & states expressed in $\sum \hat{I}_z$ basis \\
    \hline 
      \multirow{ 2}{*}{D$_1$}  & $\left| \frac{1}{2}, -\frac{1}{2} \right\rangle_1 $ & 
 $\left \lbrace 0, 0, 0, -\frac{i}{2} + \frac{1}{2 \sqrt 3 }, 0, -\frac{1}{\sqrt 3 }, \frac{i}{2} + \frac{1}{2 \sqrt3}, 0 \right \rbrace $
 \\
      & $ \left| \frac{1}{2}, +\frac{1}{2} \right\rangle_1 $ & 
  $\left \lbrace 0, -\frac{i}{2} - \frac{1}{2 \sqrt 3 }, \frac{1}{\sqrt 3}, 0, \frac{i}{2} - \frac{1}{2 \sqrt 3}, 0, 0, 0 \right \rbrace $ \\ \hline
  
     \multirow{ 2}{*}{D$_2$}  & $\left| \frac{1}{2}, -\frac{1}{2} \right\rangle_2 $ & 
$\left \lbrace 0, 0, 0, -\frac{i}{2} + \frac{1}{2 \sqrt 3 }, 0, 
  \frac{i}{2} + \frac{1}{2 \sqrt 3  }, -\frac{1}{ \sqrt 3}, 0 \right \rbrace $ \\
       & $\left| \frac{1}{2}, +\frac{1}{2} \right\rangle_2 $ & 
 $\left \lbrace 0, \frac{1}{\sqrt 3}, -\frac{i}{2} - \frac{1}{2 \sqrt 3 }, 0, \frac{i}{2} - \frac{1}{2 \sqrt 3 }, 0, 0, 0 \right \rbrace $ \\ \hline
     \multirow{ 4}{*}{Q}  & $\left| \frac{3}{2}, -\frac{3}{2} \right\rangle $ & 
$\left \lbrace 0, 0, 0, 0, 0, 0, 0, 1 \right \rbrace $ \\
       & $\left| \frac{3}{2}, -\frac{1}{2} \right\rangle $ & 
$\left \lbrace  0, 0, 0, \frac{1}{\sqrt 3}, 0, \frac{1}{ \sqrt 3 }, \frac{1}{\sqrt 3 }, 0     \right \rbrace $ \\
       & $\left| \frac{3}{2}, +\frac{1}{2} \right\rangle $ & 
$\left \lbrace  0, \frac{1}{ \sqrt 3  }, \frac{1 }{\sqrt 3 }, 0, \frac{1 }{\sqrt 3 }, 0, 0, 0    \right \rbrace $ \\
       & $\left| \frac{3}{2}, +\frac{3}{2} \right\rangle $ & 
$\left \lbrace  1, 0, 0, 0, 0, 0, 0, 0    \right \rbrace $ \\ \hline 
    \end{tabular}
\end{table}

The g-factor enhancement factor can be written as
\begin{equation}
    \gamma = \frac{g_{\text{eff}}} {g_{\text{N}}} = 2 \frac{g_e \mu_{\text{B}}} {g_{^{15}\text{N}} \mu_{\text{N}}} \left \langle f | \hat{S}_x | i\right\rangle + 2  \left \langle f \left | \sum_{j=1}^3 \hat{I}_x^{(j)} \right| i\right\rangle , \label{sieq:gamma}
\end{equation}
where the initial (final) states $\left|i \right\rangle$ ($\left|f \right\rangle$)  in \emph{leading order} is equal to $\left|0, \mathcal{I}, m_{\mathcal{I}} \right\rangle$ ($\left|-1, \mathcal{I}, m_{\mathcal{I}} \right\rangle$), where $\left|\mathcal{I}, m_{\mathcal{I}} \right\rangle$ refer to the coupled nuclear spin basis state listed in Table~\ref{sitab:states}. The second term on the right-hand side of Eq.~\ref{sieq:gamma} may take the value of 0, 1, $\sqrt{3}$, and 2 depending on the combination of the initial and final states of the nuclear doublet and the quartet subspaces considered. Since the first term on the right-hand side of Eq.~(\ref{sieq:gamma}) gives a value in the order of 10 - 1000 in the considered magnetic field interval, we neglect the second term, thus
\begin{equation}
    \gamma \approx  2 \frac{g_e \mu_{\text{B}}} {g_{^{15}\text{N}} \mu_{\text{N}}} \left \langle f | \hat{S}_x | i\right\rangle  . \label{sieq:gamma2}
\end{equation}

Non-zero matrix elements in Eq.~\ref{sieq:gamma2} are secured by the mixing of the electron and nuclear spin states. In fact, the initial and final states can be approximated as
\begin{equation}
    \left|i \right\rangle \approx \sqrt{1-\alpha^2} \left|-1, \mathcal{I}_1, m_{\mathcal{I}_1} \right\rangle + \alpha  \left|0, \mathcal{I}_2, m_{\mathcal{I}_2} \right\rangle
\end{equation}
\begin{equation}
    \left|f \right\rangle \approx \sqrt{1-\beta^2} \left|-1, \mathcal{I}_2, m_{\mathcal{I}_2} \right\rangle + \beta  \left|0, \mathcal{I}_1, m_{\mathcal{I}_1} \right\rangle .
\end{equation}
Here, $\mathcal{I}_1$ and $\mathcal{I}_2$ may refer either to the same or to two different subspaces, while  $ \Delta m_{1,2}= \left| m_{\mathcal{I}_1} - m_{\mathcal{I}_2}  \right| = 1$. Calculating the matrix element of $\hat{S}_x$ and taking the absolute value, the enhancement factor can be written as
\begin{equation}
    \gamma \approx  \sqrt{2} \frac{g_e \mu_{\text{B}}} {g_{^{15}\text{N}} \mu_{\text{N}}} \left | \alpha + \beta \right |  . \label{sieq:gamma3}
\end{equation}
The degree of mixing of the $\alpha$ and $\beta$ is state-specific, and can be obtained from  first-order perturbation theory. We assume that the spin mixing originates from the hyperfine interaction only, thus 
\begin{equation}
    \alpha \approx  \frac{  \mathcal{H} \left(-1, \mathcal{I}_1, m_{\mathcal{I}_1} ; 0, \mathcal{I}_2, m_{\mathcal{I}_2}\right) }{\left| D - g_e \mu_\text{B}B_z \right|}  =  \frac{\left\langle -1, \mathcal{I}_1, m_{\mathcal{I}_1} \right | H_\text{hyperfine}  \left|0, \mathcal{I}_2, m_{\mathcal{I}_2} \right\rangle  }{\left| D - g_e \mu_\text{B}B_z \right|}
\end{equation}
and 
\begin{equation}
    \beta \approx   \frac{  \mathcal{H} \left( -1, \mathcal{I}_2, m_{\mathcal{I}_2} ; 0, \mathcal{I}_1, m_{\mathcal{I}_1} \right) }{\left| D - g_e \mu_\text{B}B_z \right|}  =   \frac{\left\langle -1, \mathcal{I}_2, m_{\mathcal{I}_2} \right | H_\text{hyperfine}  \left|0, \mathcal{I}_1, m_{\mathcal{I}_1}  \right\rangle  }{\left| D - g_e \mu_\text{B}B_z \right|}
\end{equation}
We note that $\alpha$ and $\beta$ can be complex numbers at this point and that the mixing characterized by $\alpha$ and $\beta$ corresponds to two different nuclear hyperfine processes, namely the $\hat{S}_{\pm}\hat{I}_{\mp}$ and $\hat{S}_{\pm}\hat{I}_{\pm}$, thus $\alpha \neq \beta$ in general. Combining this with Eq.~\ref{sieq:gamma3}, we obtain
\begin{equation}
    \gamma \approx  \sqrt{2} \frac{g_e \mu_{\text{B}}} {g_{^{15}\text{N}} \mu_{\text{N}}}  \frac{\left | \delta\mathcal{H} \left( 0,-1; \ \mathcal{I}_1,\mathcal{I}_2; \ m_{\mathcal{I}_1}, m_{\mathcal{I}_2}  \right) \right | }{\left| D - g_e \mu_\text{B}B_z \right|} , \label{sieq:gamma4}
\end{equation}
where $ \delta\mathcal{H} \left( 0,-1; \ \mathcal{I}_1,\mathcal{I}_2; \ m_{\mathcal{I}_1}, m_{\mathcal{I}_2}  \right) = \mathcal{H} \left(-1, \mathcal{I}_1, m_{\mathcal{I}_1} ; 0, \mathcal{I}_2, m_{\mathcal{I}_2}\right) + \mathcal{H} \left( -1, \mathcal{I}_2, m_{\mathcal{I}_2} ; 0, \mathcal{I}_1, m_{\mathcal{I}_1} \right) $. 

\begin{table}[]
    \centering
    \caption{Nonzero matrix elements of the hyperfine Hamiltonian, and their approximate values. }
    \begin{tabular}{c|c|c}
        $\mathcal{I}_1,\mathcal{I}_2; \ m_{\mathcal{I}_1}, m_{\mathcal{I}_2}$ & $\delta\mathcal{H} $ & value (MHz) \\ \hline
        
        $Q, Q ; -\frac{1}{2}, +\frac{1}{2}$  & $\frac{\sqrt{2} \left( A_{xx}^{(1)} + A_{xx}^{(2)} + A_{xx}^{(3)} \right)}{3} $ & $129.3$ \\
        
         $Q, Q  ; \pm \frac{1}{2}, \pm\frac{3}{2}$  & $\frac{A_{xx}^{(1)} + A_{xx}^{(2)} + A_{xx}^{(3)}}{\sqrt{6}}$ & $112.0$ \\

         $D_x, D_x ; - \frac{1}{2}, +\frac{1}{2}$  & $\frac{A_{xx}^{(1)} + A_{xx}^{(2)} + A_{xx}^{(3)}}{3 \sqrt{2}}$ & $ 64.7 $ \\

        $D_2, Q ; - \frac{1}{2}, -\frac{3}{2}$  & $\frac{\sqrt 3  A_{xx}^{(1)} + \sqrt 3  A_{xx}^{(2)} - 2 \sqrt 3  A_{xx}^{(3)} - 3 A_{xy}^{(1)} - i \sqrt 3  A_{xy}^{(1)} + 3 A_{xy}^{(2)} - i \sqrt 3  A_{xy}^{(2)} + 2 i \sqrt 3 A_{xy}^{(3)} + 3 i A_{yy}^{(1)} - 3 i A_{yy}^{(2)}}{6 \sqrt{2}}$ & $  73.9 $ \\

        $D_1, Q ; + \frac{1}{2}, +\frac{3}{2}$  & $\frac{\sqrt 3  A_{xx}^{(1)} - 2 \sqrt 3  A_{xx}^{(2)} + \sqrt 3 A_{xx}^{(3)} + 3 A_{xy}^{(1)} -  i \sqrt 3  A_{xy}^{(1)} + 2 i \sqrt 3  A_{xy}^{(2)} - 3 A_{xy}^{(3)} - i \sqrt 3  A_{xy}^{(3)} -  3 i A_{yy}^{(1)} + 3 i A_{yy}^{(3)}}{6 \sqrt{2}}$ & $  73.9 $ \\

         $D_2, D_1 ; - \frac{1}{2}, +\frac{1}{2}$  & $\frac{-2 A_{xx}^{(1)} + A_{xx}^{(2)} + A_{xx}^{(3)} + 2 I A_{xy}^{(1)} - I A_{xy}^{(2)} - \sqrt 3  A_{xy}^{(2)} - i A_{xy}^{(3)} + \sqrt 3  A_{xy}^{(3)} + i \sqrt 3  A_{yy}^{(2)} - i \sqrt 3  A_{yy}^{(3)}}{3 \sqrt{2}}$ & $  42.7 $ \\

        $Q, D_2 ; - \frac{1}{2}, +\frac{1}{2}$  & $\frac{ \sqrt 3  A_{xx}^{(1)} + \sqrt 3  A_{xx}^{(2)} - 2 \sqrt 3  A_{xx}^{(3)} - 3 A_{xy}^{(1)} -  i \sqrt 3  A_{xy}^{(1)} + 3 A_{xy}^{(2)} - i \sqrt 3 A_{xy}^{(2)} + 2 i \sqrt 3  A_{xy}^{(3)} +  3 i A_{yy}^{(1)} - 3 i A_{yy}^{(2)}}{6 \sqrt{6}}$ & $  21.3 $ \\

        $D_1, Q ; - \frac{1}{2}, +\frac{1}{2}$  & $\frac{ \sqrt 3 * A_{xx}^{(1)} - 2 \sqrt 3  A_{xx}^{(2)} + \sqrt 3  A_{xx}^{(3)} + 3 A_{xy}^{(1)} -  i \sqrt 3  A_{xy}^{(1)} + 2 i \sqrt 3  A_{xy}^{(2)} - 3 A_{xy}^{(3)} - i \sqrt 3  A_{xy}^{(3)} - 3 i A_{yy}^{(1)} + 3 i A_{yy}^{(3)}}{6 \sqrt{6}}$ & $  21.3 $ \\

    \end{tabular}
    \label{sitab:mez}
\end{table}

Table~\ref{sitab:mez} provides all the nonzero matrix elements of the hyperfine Hamiltonian term and their approximate values obtained from first principles theory. Altogether, we observe 10 non-zero terms, which all give rise to a large g-factor enhancement. The largest enhancement is obtained for transitions that occur within the quartet subspace, see the top three matrix elements in Table~\ref{sitab:mez}.

For these terms, the g-factor enhancement is
\begin{equation}
    \gamma \approx  \frac{\xi}{3} \frac{g_e \mu_{\text{B}}} {g_{^{15}\text{N}} \mu_{\text{N}}}  \frac{ \left | A_{xx}^{(1)} + A_{xx}^{(2)} + A_{xx}^{(3)}  \right |  }{\left| D - g_e \mu_\text{B}B_z \right|} , \label{sieq:gamma5}
\end{equation}
where $\xi $ is a prefactor that originates from the matrix elements of ladder operators of the nuclear spin subspaces and takes the values 2 and $\sqrt{3}$ for $+1/2 \leftrightarrow -1/2$ and $ \pm 1/2 \leftrightarrow \pm 3/2$ transitions in the quartet subspace, respectively, and the value of 1 for  the $+1/2 \leftrightarrow -1/2$ transitions in both doublet subspaces. 

In addition to the five intra-subspace transitions, we observe nonzero matrix elements with the inter-subspace transitions, see Table~\ref{sitab:mez}.  The analytical curves are plotted together with the results of exact denationalization in Fig.~3d of the main text.

\section{Supplementary Note 6 - Comparison of ENDOR spectra for $m_s=\pm1$}

\begin{figure}
    \centering
    \includegraphics
    {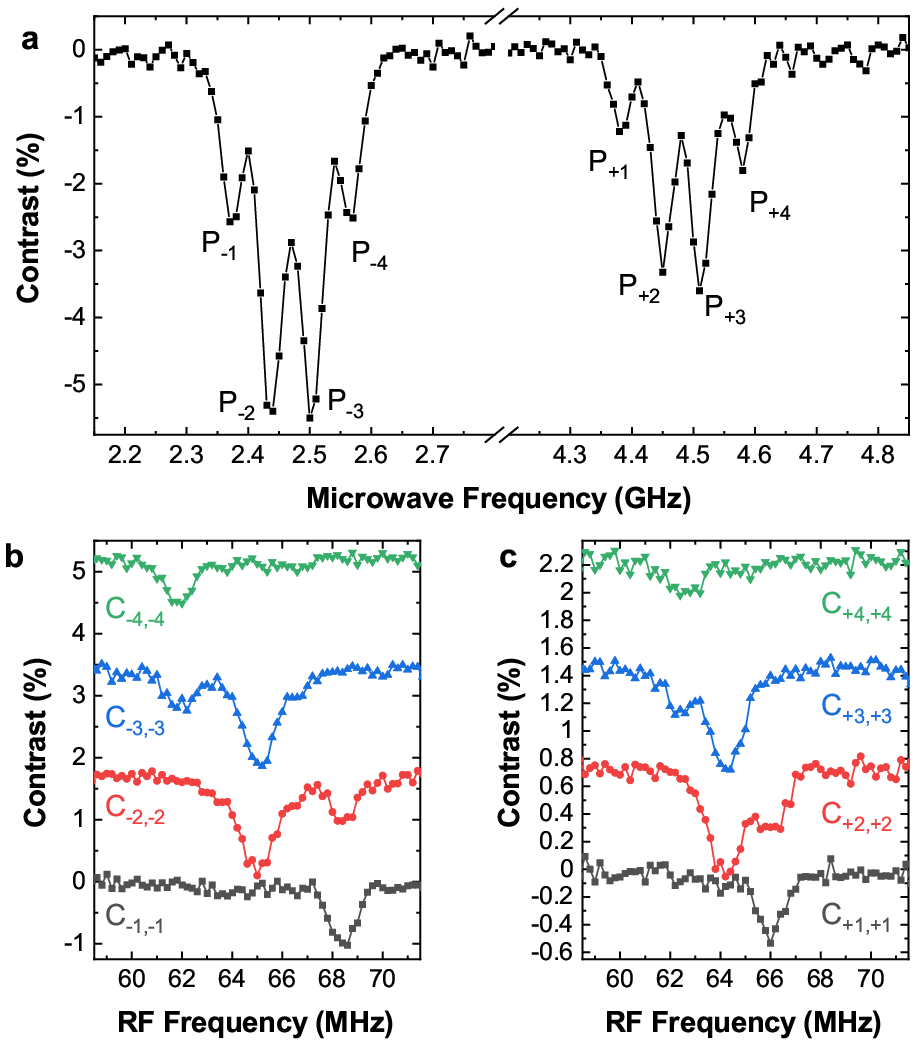}
    \caption{ Comparison of ENDOR spectra for $m_s=\pm1$ \textbf{a} Representative Rabi Oscillations at different RF power. \textbf{b} Nuclear Rabi Frequency versus RF power. \textbf{c} Nuclear Rabi Coherence Time as a function of RF Rabi Frequency.}
    \label{fig:ENDOR_35mT}
\end{figure}

In Fig. \ref{fig:ENDOR_35mT}a the ODMR spectrum for a magnetic field $B_z=35~\mathrm{mT}$ showing both the $m_s=-1$ and $m_s=+1$ transitions, with the corresponding ENDOR in Fig. \ref{fig:ENDOR_35mT}b and c, respectively.

\section{Supplementary Note 7 -  Nuclear Rabi Oscillations}

\begin{figure}
    \centering
    \includegraphics[width=\linewidth]{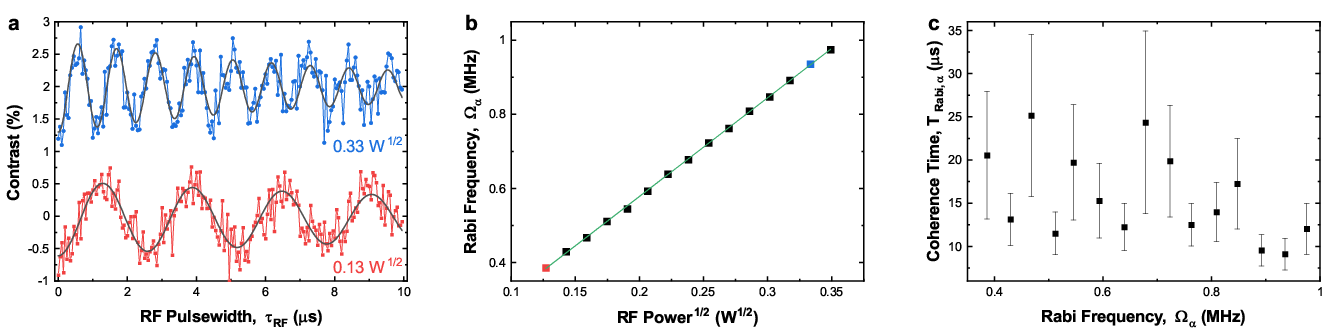}
    \caption{Experimental Nuclear Rabi Oscillations. \textbf{a} Representative Rabi Oscillations at different RF power. \textbf{b} Nuclear Rabi Frequency versus RF power. \textbf{c} Nuclear Rabi Coherence Time as a function of RF Rabi Frequency.}
    \label{fig:Nuclear_Rabi}
\end{figure}

Nuclear Rabi oscillation measurements are performed as a function of the RF peak power, as shown in Fig. \ref{fig:Nuclear_Rabi}a for the $\alpha$ nuclear transition at $B_z=212~\mathrm{mT}$. The Rabi frequency $\Omega_\alpha$ and Rabi coherence time $T_{Rabi,\alpha}$ are extracted using an exponentially damped cosine fit, $Ae^{-\tau/T_{Rabi,\alpha}}\cos(\Omega_e\tau_{RF}+\theta)$. The Rabi frequency is plotted as a function of the RF amplitude in Fig. \ref{fig:Nuclear_Rabi}b, showing the expected linear dependence. As shown in Fig. \ref{fig:Nuclear_Rabi}, the nuclear Rabi coherence time, shows a downwards trend with increasing Rabi Frequency. However, the accuracy of the fitting is limited by the range of RF pulsewidth, which is limited to $\sim10~\mathrm{\mu s}$ by the electron $T_1$ time. 

\section{Supplementary Note 8 -  Magnetic Field Alignment}

\begin{figure}
    \centering
    \includegraphics[width=0.8\linewidth]{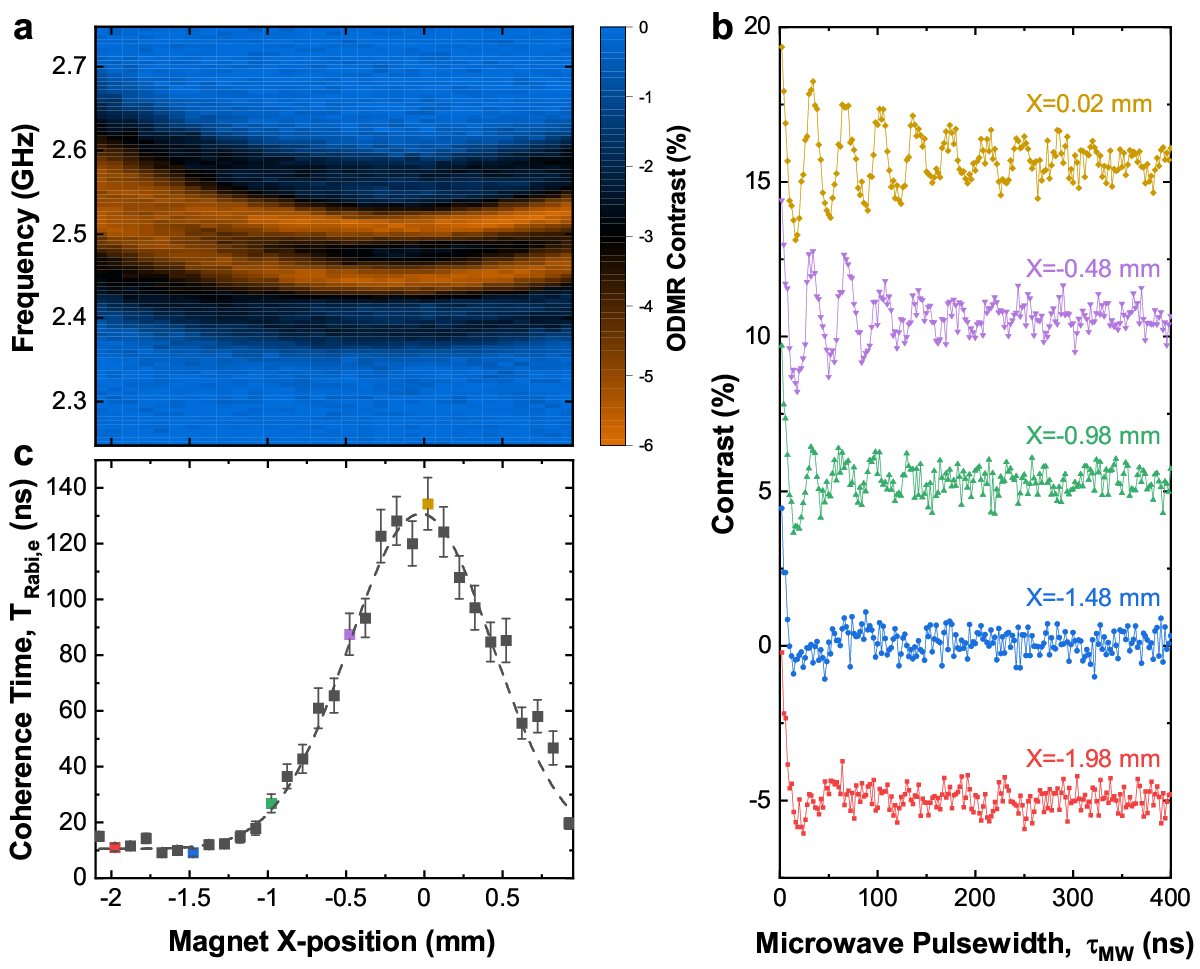}
    \caption{Alignment of the static magnetic field with the $V_B^-$ quantization axis. \textbf{a} ODMR spectra at $B_z=35~\mathrm{mT}$ \textbf{b,c} ENDOR spectra for \textbf{b} $m_s=-1$ and \textbf{b} $m_s=+1$ transitions.}
    \label{fig:B_align}
\end{figure}

The static magnetic field is applied using a cylindrical permanent magnet mounted on an XYZ-translation stage with the poles of the magnet along the Z-axis. To align the magnetic field with the quantization axis of the $V_B^-$ defects we first measure ODMR spectra as a function of the magnets X-position, as shown in Fig. \ref{fig:B_align}a. The frequencies of the four hyperfine transitions are extracted from a multiple Gaussian fit at each X-position. A Rabi-oscillation is then measured for each X-position with a microwave frequency corresponding to the $m_{Jz}=+1/2$ transition. Example measurements are shown in Fig. \ref{fig:B_align}b. Finally, the Rabi coherence time $T_{Rabi}$ is extracted at each X-position with a fit to $Ae^{-\tau/T_{Rabi,e}}\cos(\Omega_e\tau_{MW}+\theta)$. As shown in Fig. \ref{fig:B_align}c, there is a clear maxima in the coherence, similar to the case for ensembles of NV centers in diamond\cite{Stanwix_PRB_2010}, and which we fit phenomenologically with a Gaussian to find the optimal magnet position. The procedure is repeated for the magnet's Y-position. When properly aligned $T_{Rabi,e}\approx130~\mathrm{ns}$ for $\Omega_e\approx20~\mathrm{MHz}$, the conditions used in ENDOR measurements.  


\clearpage


\printbibliography